\documentclass[aps,twocolumn,superscriptaddress,amsmath,amssymb,floatfix]{revtex4-1}

\usepackage{amsfonts,hyperref}
\usepackage{graphicx}
\usepackage{amsmath}
\usepackage{amssymb}
\usepackage{mathrsfs}
\usepackage{bm} 
\usepackage{bbold} 
\usepackage{longtable}
\usepackage{outlines}
\usepackage{xcolor}

\newcommand{\bra}[1]{\mbox{$\langle #1 |$}}
\newcommand{\ket}[1]{\mbox{$| #1 \rangle$}}
\newcommand{\braket}[2]{\mbox{$\langle #1 | #2 \rangle$}}
\newcommand{\ZZ}{$Z_2$ }

\newcommand{\hsp}{\hspace{0.6cm}}

\providecommand{\e}[1]{\ensuremath{\times 10^{#1}}}

\begin{document}
\title{Detecting a \texorpdfstring{\ZZ}{Z2 }topologically ordered phase from unbiased infinite projected entangled-pair state simulations}
\author{S.P.G.~Crone}
\affiliation{Institute for Theoretical Physics and Delta Institute for
Theoretical Physics, University of Amsterdam, Science Park 904, 1098 XH
Amsterdam, The Netherlands}

\author{P.~Corboz}
\affiliation{Institute for Theoretical Physics and Delta Institute for
Theoretical Physics, University of Amsterdam, Science Park 904, 1098 XH
Amsterdam, The Netherlands}

\date{\today}

\begin{abstract}
We present an approach to identify topological order based on unbiased infinite projected entangled-pair states (iPEPS) simulations, i.e. where we do not impose a virtual symmetry on the tensors during the optimization of the tensor network ansatz. As an example we consider the ground state of the toric code model in a magnetic field  exhibiting \ZZ topological order. The optimization is done by an efficient energy minimization approach based on a summation of tensor environments to compute the gradient.
We show that the optimized tensors, when brought into the right gauge, are approximately \ZZ symmetric, and they can be fully symmetrized a posteriori to generate a stable topologically ordered state, yielding the correct topological entanglement entropy and modular S and U matrices. To compute the latter we develop a variant of the corner-transfer matrix method which is computationally more efficient than previous approaches based on the tensor renormalization group. 
\end{abstract}	

\maketitle

\section{Introduction}
Since the discovery of the fractional quantum Hall effect~\cite{tsui82}, the understanding of topologically ordered phases has been a central subject in many-body physics. These phases do not fall under the standard paradigm of Landau symmetry breaking theory and therefore cannot be characterized in terms of a local order parameter~\cite{wen1990}. They exhibit remarkable properties, including a ground state degeneracy depending on the topology of the system, emergent anyonic excitations~\cite{kitaev2003,kitaev2006}, and they are robust against local perturbations which makes them a promising platform for quantum computing~\cite{nayak08}. However, in general it has proven to be challenging to determine, starting from a microscopic Hamiltonian $\hat{H}$, whether the ground state is in a topological ordered phase.

In recent years substantial progress in studying and classifying these phases has been achieved on the basis of tensor networks. Many studies based on matrix product states (MPS) have shown that topologically ordered phases can be identified~\cite{yan2011,jiang2012,jiang2012b,depenbrock2012}, including the characterization of their emerging anyonic excitations, see e.g. Refs.~\cite{cincio2013,zaletel13,he_chiral_2014,bauer_chiral_2014,zhu_topological_2015,grushin_characterization_2015,he_kagome_2015,he_distinct_2015}. However, due to the one-dimensional nature of the MPS ansatz, these studies are limited to cylinders up to a certain width.
Projected entangled-pair states (PEPS)~\cite{verstraete2004,nishio2004,verstraete2006}, which are a  generalization of MPS to two dimensions~(2D), provide a more natural framework for the study of  2D topologically ordered systems~\cite{gu2008,schuch2010}. There exist a wide range of (non-chiral~\footnote{We note that there has also been interesting progress in the study of chiral topologically ordered states with PEPS recently~\cite{chen18,lee19}}) topologically ordered states which have an exact and simple PEPS representation, such as e.g. the ground states of the toric code model~\cite{verstraete2006}, string-net models~\cite{buerschaper09,gu2009}, or resonating valence-bond states~\cite{schuch12b}. It has been shown that the topological order is encoded locally in the PEPS tensors by respecting a certain symmetry on their virtual degrees of freedom~\cite{schuch2010}, depending on the type of topological order. The characterization of these virtual symmetries and their associated topologically ordered phases has been under active development in recent years~\cite{schuch2013,Buerschaper2014,sahinoglu2014,haegeman2015,duivenvoorden17,bultinck2017,iqbal18}.

However, it has been shown that already a weak violation of the virtual symmetry destroys the associated topological order~\cite{chen2010,shukla2018}. Thus, in practical calculations, when performing an optimization starting from random initial tensors, one may expect that already small numerical errors in the optimization of the tensor network ansatz will result in tensors which are not perfectly symmetric, and thus it seems challenging to correctly identify a topological ordered phase. A way to circumvent this problem is to impose the  virtual symmetry on the tensors during the optimization~\cite{he2014a}, but this requires knowledge of the virtual symmetry beforehand. Without a priori knowledge of the ground state of a given Hamiltonian, one would need to run many simulations, starting from tensors with different virtual symmetries, in order to identify the true ground state (given by the state with lowest variational energy). Since the optimization of the tensors is the computationally most expensive part in a tensor network calculation, it would be  desirable to be able to start from unbiased simulations (i.e. without imposing a virtual symmetry), and to identify the topologically order  a posteriori. However, due to the sensitivity to perturbations, it has so far been unclear whether this is actually possible in practice (up to very recently \cite{francuz19}, see comment below). 

In this work we demonstrate that the study and identification of topological ordered phases with infinite PEPS (iPEPS)~\cite{jordan2008} is indeed feasible even without imposing the virtual symmetry on the tensors during the optimization. As an example we consider the toric code model with an external magnetic field, where the tensors are known to exhibit a \ZZ virtual symmetry in the topologically ordered phase.
We show that the  resulting tensors (after a suitable gauge change) are approximately \ZZ symmetric, and  that they can be fully symmetrized after the optimization. The resulting state exhibits the relevant features of the topologically ordered state, including the correct topological entanglement entropy and modular S and U matrices characterizing the mutual and self-statistics of the emergent anyonic excitations. 

Following ideas from Ref.~\onlinecite{zhang2012}, the modular matrices are obtained from the computation of wave function overlaps of a complete set of ground states on a torus with minimum entanglement entropy, where each state has a well defined anyonic flux through the torus. This approach has already successfully been applied based on  matrix product states on cylinders~\cite{cincio2013}, and also with iPEPS using the  tensor renormalization group (TRG) method~\cite{he2014a,huang16,mei17,chen18b, gauthe19}. In this work we introduce another scheme based on the  corner transfer matrix (CTM) renormalization group method~\cite{nishino1996} which is more efficient than TRG to compute wave function overlaps.

We note that during completion of this work, a different approach to study topological order based on iPEPS without imposing a virtual symmetry was presented in Ref.~\cite{francuz19}. Instead of recovering the virtual symmetry of the local tensors, the approach in Ref.~\cite{francuz19} is based on projectors onto the ground states with different anyonic fluxes represented by matrix product operators (MPO), which are found by an optimization procedure.

A technical challenge when simulating the square lattice toric code model with iPEPS is that the Hamiltonian consists of four-body operators.
While previous energy minimization algorithms~\cite{corboz2016,vanderstraeten2016} are in principle not  restricted to nearest-neighbor models~\cite{niesen18}, their generalization to more complicated Hamiltonians is  rather tedious and computationally expensive. In the present work we have  developed an alternative energy minimization algorithm which, besides simple nearest-neighbor terms, can treat more general Hamiltonian operators in a simpler and more efficient way. (We note that  recently another scheme based on automatic differentiation~\cite{liao2019} has been introduced with a similar computational cost.) 

This paper is organized as follows. In Sec.~\ref{chap:TC}, the toric code model is introduced as an example of a system with \ZZ topologically order, together with basic notions and concepts including the topological entanglement entropy (TEE), minimum entropy states (MES), and modular matrices. Then, an introduction to iPEPS and the standard corner transfer matrix (CTM) algorithm is given in Sec.~\ref{Sec:iPEPSansatz} and ~\ref{Sec:CTM}. In Sec.~\ref{sec:GradientOpt} we explain the gradient-based optimization scheme developed in this work. 
In Sec.~\ref{chap:Framework}, we present our approach to identify a topologically ordered phase with iPEPS, with the \ZZ topological order as an example. This includes the scheme to recover the \ZZ virtual symmetry in the tensors explained in Sec.~\ref{Chap:RestorSym}, and the extension of the CTM method to compute wave function overlaps of the MES to determine the modular matrices in Sec.~\ref{Chap:Calculation_CTM_matrices}. In Sec.~\ref{chap:Results} we present results for the toric code model obtained with our approach, and end with our conclusions in Sec.~\ref{Sec:summary}.


\section{The toric code model}
\label{chap:TC}

\begin{figure}[tb]
\begin{center}
	\includegraphics[width=\columnwidth]{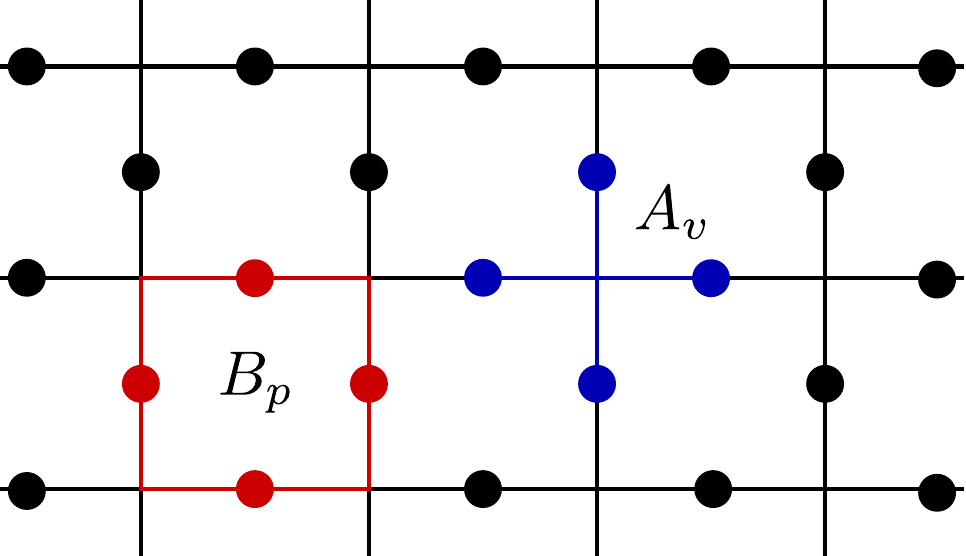}
\caption{The square lattice for the toric code Hamiltonian considered in this work. The black dots on the edges represent spin-$\frac{1}{2}$ particles. The spins involved in the action of a single $A_v$ or $B_p$ operation is shown in blue and red, respectively. }
\label{Fig:GridTC}
\end{center}
\end{figure}

In this work we consider the square lattice toric code model \cite{kitaev2003} as a simple example of a system exhibiting a \ZZ topologically ordered ground state. The model consists of spin-$\frac{1}{2}$ degrees of freedom placed on the edges of the lattice, as shown in Fig.~\ref{Fig:GridTC}. The Hamiltonian is given by 
\begin{equation}
	\hat{H}_{\mathrm{TC}} = -\sum_v A_v - \sum_p B_p 
	\label{Eq:TC_Hamiltonian}	
\end{equation}
where $v$ denotes the vertices of the lattice, $p$ the plaquettes. The operator $A_v = \prod_{i \in v} \sigma^{z}_{i}$ is given as a product of Pauli matrices $\sigma^{z}_{i}$ acting on spin $i$ adjacent to a vertex~$v$, and similarly $B_p = \prod_{i \in p} \sigma^{x}_{i}$ is a product of Pauli matrices $\sigma^{x}_{i}$ acting on the spins on a plaquette $p$, as shown in Fig.~\ref{Fig:GridTC}. 

The model is exactly solvable, where the solution can best be seen in the $\sigma_{z}$ basis where the spin up configuration is associated with the presence of a line going through the spin.  All the plaquette and vertex terms commute with each other, so the ground state configurations are the states where all plaquette and vertex terms have eigenvalue +1, i.e. $A_v\ket{\Psi} = B_p \ket{\Psi} = \ket{\Psi}$. On a vertex this requires that the number of up spins (and down spins) is even which in the line representation  means that if a line enters a vertex it also has to leave the vertex. Because lines cannot end at a vertex, the ground state can only contain configurations of closed loops of lines of up spins. The plaquette term in this basis flips all the spins around a plaquette, which allows this term to locally connect states with different closed loop configurations. The ground state, being an eigenstate of both terms simultaneously, therefore has to consist of an equally weighted superposition of all the possible closed loop configurations.

\begin{figure}[tb]
\begin{center}
	\includegraphics[width=\columnwidth]{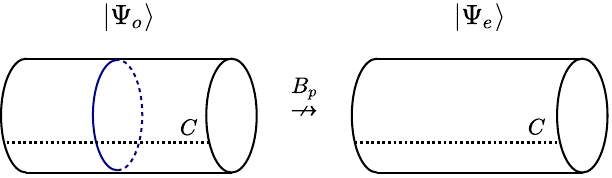}
	\caption{The ground state $\ket{\Psi_o}$ cannot be turned into $\ket{\Psi_e}$ under the action of the Hamiltonian, resulting in two distinct degenerate ground states. The two ground states can be distinguished by  the non-local operator $W^{(z)}(C) = \prod_{i \in C} \sigma^z_i$ along the line $C$ which measures the parity of the number of loops cutting the line $C$. }
	\label{Fig:CylCalc}
\end{center}
\end{figure}

The ground state exhibits a degeneracy which depends on the topology of the lattice. This because the plaquette term can only transform a closed loop configuration locally, but it is not able to remove a non-contractible loop which winds around periodic boundaries. For example, when considering the model on a cylinder geometry, a single loop going around the cylinder cannot be removed by the plaquette operator (see Fig.~\ref{Fig:CylCalc}). This results in two different ground state sectors which can be labeled by the parity of the number of loops winding around the cylinder. The ground state manifold is spanned by  $\{\ket{\Psi_e},\ket{\Psi_o}\}$ where $e$ and $o$ denote the even and odd sectors, respectively. On a torus the ground state degeneracy is fourfold, and the ground states can be labelled 
as $\{\ket{\Psi_{ee}},\ket{\Psi_{eo}},\ket{\Psi_{oe}},\ket{\Psi_{oo}} \}$ according to the parity in horizontal and vertical direction.

Besides the ground state, also the elementary anyonic excitations in the toric code model are well-understood. They correspond to a violation of a vertex term (electric excitation) or a plaquette term (magnetic excitation) and they always occur in pairs. Acting with a $\sigma^x_i$ operator on the ground state violates both vertex terms including site $i$, i.e. it creates two electric (e) particles on the adjacent vertices. More generally, the path operator $W^{(x)}(\gamma) = \prod_{i \in \gamma} \sigma^x_i$ where $\gamma$ is an open path on the lattice creates two electric particles at its endpoints. Similarly, a $\sigma^z_i$ operator on the ground state violates both plaquette terms including site $i$, corresponding to a creation of two magnetic particles located on the two adjacent plaquettes, and the path operator $W^{(z)}(\bar \gamma) = \prod_{i \in \bar \gamma} \sigma^z_i$ with $\bar \gamma$ an open path on the dual lattice creates a pair of particles at the endpoints of this path. While both excitations have trivial (bosonic) self-statistics, they exhibit non-trivial mutual statistics since a phase factor -1 is acquired upon braiding an e and an m particle. One can further identify the composite particle~$\epsilon$ corresponding to a combination of an e and m particle, which has fermionic self-statistics, and the trivial identity particle~$\mathbb{1}$.

\subsection{Topological entanglement entropy and minimal entropy states}
\label{Sec:TEE}
Topologically ordered states are known to exhibit a universal correction $\gamma$ to the area-law of entanglement~\cite{levin2006,kitaev2006a},  $S_A(L) \sim \alpha L - \gamma$, where $S(L)$ is the entanglement entropy  between a disk-shaped region $A$ with a smooth boundary of length $L$ and the rest of the system~\footnote{We note that the value is independent of  the Renyi entropy index}. The universal constant $\gamma$ is called the topological entanglement entropy (TEE) and is equal to $\log \mathcal{D}$, where $\mathcal{D} = \sum_k \sqrt{(d_k^2)}$ is the total quantum dimension and $d_k$  the quantum dimensions of the $k$th particle type of the underlying topological theory. For abelian anyons, $d_k = 1$  $\forall k$, and therefore $\gamma = \log{2}$ for the toric code ground state~\cite{hamma2005,kitaev2006a}. %
A finite TEE is a characteristic feature of a topologically ordered phase and can thus be used to detect such a phase~\footnote{We note that the TEE does not uniquely identify a topologically ordered phase, since two different topologically ordered phases can have the same value for $\gamma$.}. Special care must  be taken, however, if the region A has a nontrivial topology. If the boundary of $A$ is non-contractible, the value obtained for $\gamma$ depends on the specific ground state, and only the so-called minimum entropy states (MES) yield the maximal, universal value for $\gamma$~\cite{zhang2012}. 
 
Consider a torus cut into two cylindric (i.e. non-contractible) regions A and B. The MES correspond to ground states with a well defined anyonic flux through the entanglement cut between A and B~\cite{zhang2012}. For the toric code the four different MES for this bipartition can be identified with the fluxes of anyons $\{\mathbb{1}, e, m, \epsilon\}$,  corresponding to eigenstates of the loop operators $W^{(z)}(C)$ and $W^{(x)}(C)$ acting along the cut $C$ which detect the presence of an electric and magnetic flux, respectively. One can easily show~\cite{zhang2012}  that with  respect to a cut in vertical direction the MES are given by
\begin{align}
	\begin{split}
	\label{Eq:MESs}
	\ket{\Psi_{\mathbb{1}/m}} &=   \frac{1}{\sqrt{2}}[\ket{\Psi_{ee}} \pm \ket{\Psi_{eo}}], \\
	\ket{\Psi_{e/\epsilon}} &= \frac{1}{\sqrt{2}}[\ket{\Psi_{oe}} \pm \ket{\Psi_{oo}}]. \\
	\end{split}
\end{align}
Any of the MES will yield the universal constant $\gamma = \log{2}$  when computing the TEE of region A. We will show in Sec.~\ref{Chap:Calculation_CTM_matrices} how we can obtain the MES based on iPEPS.

\subsection{Modular S and U matrices}
Because the MES describe states with different anyonic fluxes, they can be used to compute the modular $S$  and $U$ matrices which characterize the non-trivial braiding and self-statistics of the anyonic excitations, as shown in Ref.~\onlinecite{zhang2012}. For  abelian anyons, the $S_{ij}$ matrix describes the phase a particle $i$ obtains when encircling particle $j$ (divided by the total quantum dimension). The $U_{ij}$ matrix, which is diagonal, describes the phase a particle $i$ obtains when exchanged with another particle of type~$i$. 
 
The $S$-matrix, which on a square geometry acts as a $\pi/2$ rotation on the MES basis, can be calculated as~\cite{wen1990}
\begin{equation}
	S_{ij} = \frac{1}{D} \braket{\Psi_{i}^{\hat{y}}}{\Psi_{j}^{\hat{x}}}
	\label{Eq:Smatrix}
\end{equation}
where $D$ is the total quantum dimension, and the states $\ket{\Psi_{i}^{\hat{y}}}$  and $\ket{\Psi_{j}^{\hat{x}}}$ denote the MES with anyonic flux $i$ in $\hat x$-direction and anyonic flux $j$ in the perpendicular $\hat{y}$ direction on the  torus, respectively. 

 The $U$ matrix describes the action of a Dehn twist on the torus which can be viewed as  as cutting  the torus  along  the $\hat{y}$ direction to create a cylinder, rotating one of the cuts by $2\pi$, and glueing the cuts back together to get back a torus geometry. If an anyonic flux is going perpendicularly through the cut, this one gets wrapped around the $\hat{y}$ direction of the cut. Therefore, the $U$ matrix can be viewed as an operation which adds the flux present in $\hat{x}$-direction to the flux along the $\hat{y}$-direction (cf. also Ref.~\onlinecite{he2014a}). Thus, on a state with parity $p_x$ and $p_y$ in the x- and y-direction, respectively, one obtains $\hat U \ket{\Psi_{p_x p_y}} = \ket{\Psi_{p_x (p_y p_x)}}$, and thus one finds for the MES of the toric code
\begin{align}
	\begin{split}
	\hat{U} \ket{\Psi_{\mathbb{1}/m}} &= \frac{1}{\sqrt{2}}[\ket{\Psi_{ee}} \pm \ket{\Psi_{eo}} =  \ket{\Psi_{\mathbb{1}/m}}  \\
	\hat{U} \ket{\Psi_{e/\epsilon}} &= \pm \frac{1}{\sqrt{2}}[\ket{\Psi_{oe}} \pm \ket{\Psi_{oo}}] = \pm \ket{\Psi_{e/\epsilon}}	
	\label{Eq:Umatrix}
	\end{split}
\end{align}

\subsection{Toric code in an external magnetic field}
Besides the standard toric code model, in this work we also consider the model in a  magnetic field,
\begin{equation}
	\hat{H} = J \hat{H}_{\mathrm{TC}} - h_z \sum_{i}  \sigma_i^{z} - h_x \sum_{i}  \sigma_i^{x}
	\label{Eq:TC_Hamiltonianh}	
\end{equation}
where $h_z$ and $h_x$ are the magnetic field strengths in $z$ and $x$ direction, respectively, and we will set $J=1/2$ in the following. This model is no longer exactly solvable, but has been studied in previous works by series expansions~\cite{vidal2009a,dusuel2011}, Monte-Carlo methods~\cite{wu2012} and iPEPS~\cite{dusuel2011,vanderstraeten17}. The topological phase extends to a finite value of the magnetic field where a phase transition occurs towards a magnetically ordered phase. When the magnetic field is only  along either the $h_z$ or $h_x$ direction, the model can be mapped to a 2D transverse field Ising model~\cite{trebst2007}, with a second order transition occurring at $h_z(h_x)= 0.164237(2)$~\cite{wu2012}. When the field is applied in the  $h_x=h_z$ direction, the model can be mapped onto the 3D classical \ZZ gauge Higgs model~\cite{tupitsyn2010} with a second order transition at $h_x=h_z=0.170(1)$~\cite{wu2012}.


\section{iPEPS}
\label{chap:iPEPSMethod}

\subsection{iPEPS ansatz}
\label{Sec:iPEPSansatz}

\begin{figure}[tb]
\begin{center}
\includegraphics[width=\columnwidth]{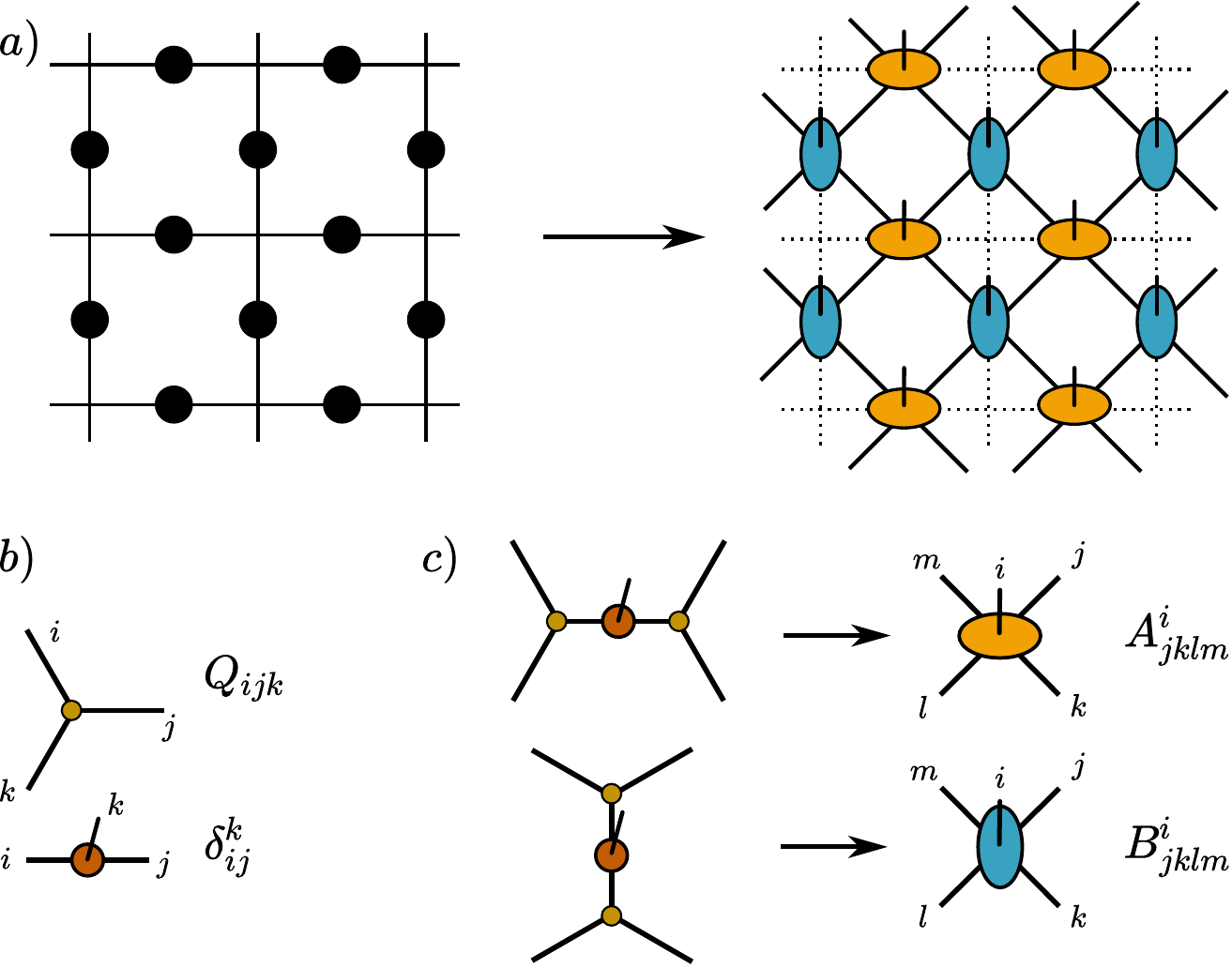}
\caption{(a) Mapping the two dimensional lattice to an iPEPS representation with a checkerboard pattern of $A$ (orange) and $B$ (blue) tensors. These tensors are related by a 90$^\circ$ rotation. (b),(c) Tensors used for the construction of the exact ground state of the toric code model.}
\label{Fig:MapLatticeTN}
\end{center}
\end{figure}

An infinite projected entangled-pair state (iPEPS)~\cite{verstraete2004,nishio2004,jordan2008} is a tensor network ansatz which can systematically approximate  ground states of two-dimensional lattice models in the thermodynamic limit. The ansatz exploits the area-law of entanglement of gapped local Hamiltonians~\cite{eisert2010}, which a PEPS reproduces by construction. On a square lattice, an iPEPS consists of a periodically repeated unit cell made up of tensors with five indices (legs), as shown in Fig.~\ref{Fig:MapLatticeTN}.  Each tensor has a single physical leg representing the local Hilbert space of one or more lattice sites, and four auxiliary legs which connect to the neighboring tensors on square lattice. The accuracy of the ansatz is systematically controlled by the dimension of the auxiliary indices called the bond dimension $D$. 

In this work, the lattice of the toric code model is mapped onto a tensor network with a unit cell consisting of two tensors $A^{i}_{jklm}$ and $B^{i}_{jklm}$ arranged in a chequerboard pattern, see   Fig.~\ref{Fig:MapLatticeTN}(a). The $A$ ($B$) tensors represent the spins on the horizontal (vertical) bonds. Due to the symmetry of the lattice we choose the  tensors to exhibit mirror symmetries, $A^{i}_{jklm} = A^{i}_{kjml}$ and $A^{i}_{jklm} = A^{i}_{mlkj}$, where the two tensors are related as $B^{i}_{jklm} = A^{i}_{klmj}$. 

The ground state of the toric code (without magnetic field) can be exactly represented with $D=2$ tensors as defined in Fig.~\ref{Fig:MapLatticeTN}(b), where the $Q$ and $\delta$ tensors are zero except for the elements
\begin{eqnarray}
	Q_{111}&=&Q_{221}=Q_{212}=Q_{122}=1, \\
	\delta^1_{11}&=&\delta^2_{22}=1.
\end{eqnarray}
From the figure one can see that the two tensors $A$ and $B$ are related by a rotation of 90 degrees.~\footnote{Note that the representation is not unique and other representations were used e.g. in Refs.~\onlinecite{verstraete2006,gu2008}.}

In the following we discuss the main ingredients of the iPEPS algorithm, including the contraction of the tensor network  and the optimization of the tensors, i.e. finding the best variational parameters of the tensors which minimize the variational energy $E = \bra{\Psi} \hat{H} \ket{\Psi}/ \braket{\Psi}{\Psi}$ where $\ket{\Psi}$ is the iPEPS wave function. 

\subsection{Corner Transfer Matrix algorithm}
\label{Sec:CTM}

\begin{figure}[tb]
\begin{center}
	\includegraphics[width=\columnwidth]{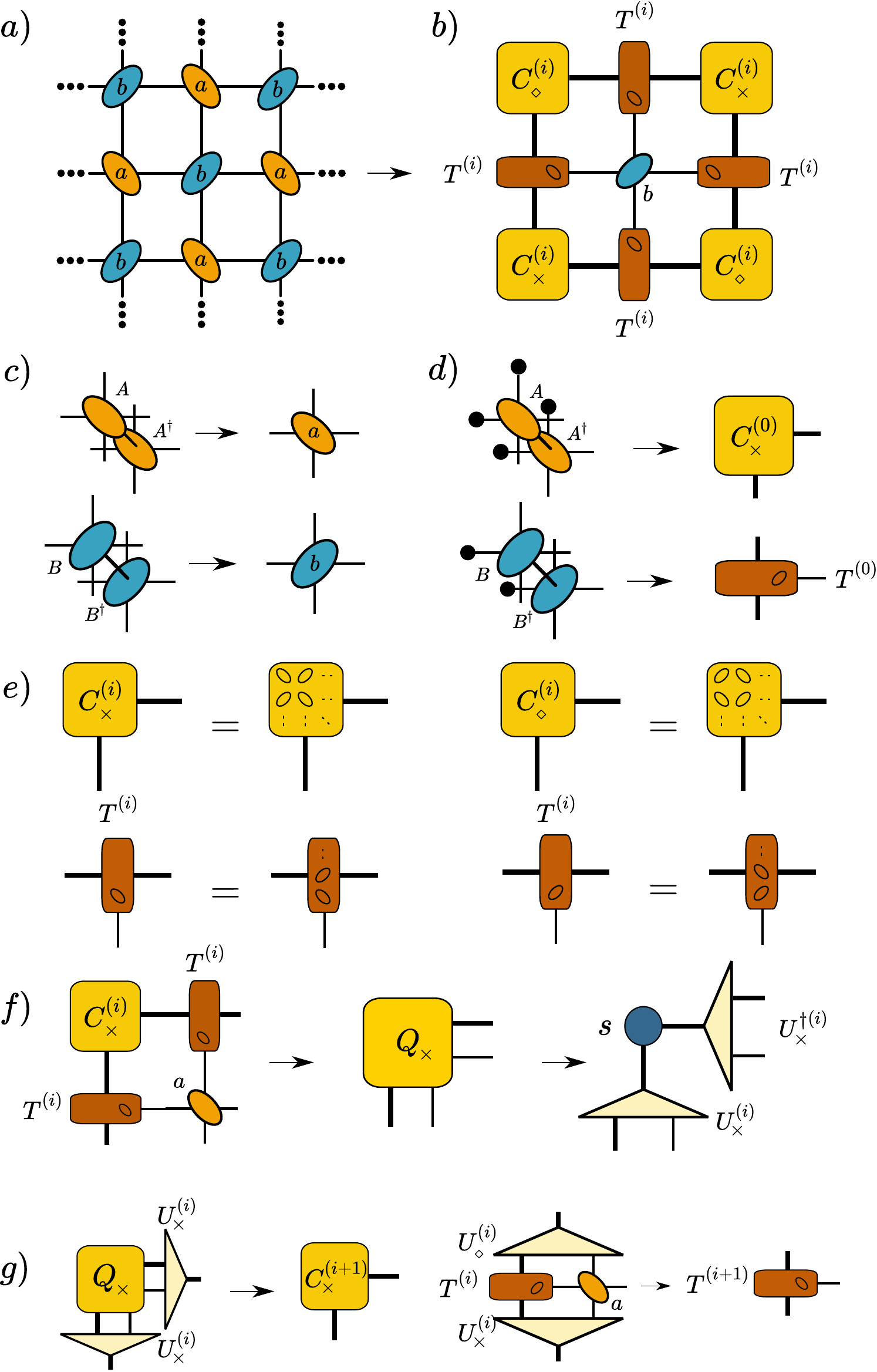}
\caption{Tensor network diagrams describing the (symmetric) CTM algorithm used in this work. $(a)$ Infinite tensor network consisting of the checkerboard pattern of tensors $a$ and $b$, constructed by combining the iPEPS tensors $A$ with $A^{\dagger}$ and  $B$ with $B^{\dagger}$, respectively, as shown in~(c). The infinite tensor network surrounding a bulk tensor $b$ is effectively represented by an environment consisting of four corner tensors $C_{\times}$ / $C_{\diamond}$  and four edge tensors $T$ as depicted in (b). The thin and thick lines have a dimension $D^2$ and $\chi$, respectively.  (d)~The initial boundary tensors are constructed from the iPEPS bulk tensors by projecting the legs in both the bra and ket layer onto the first element, depicted by the black dots. (e) Definition of the two different corner tensors and orientation of the $T$ tensor. (f-g) Growth and renormalization step in the CTM algorithm (cf. text).}
\label{Fig:CTMsteps}
\end{center}
\end{figure}

To compute an expectation value of an observable with iPEPS, the corresponding operator is placed between the bra and ket layer and the resulting two-dimensional network is contracted. The contraction of the two-dimensional network cannot be done in an exact way but only approximately. In this work we use the Corner Transfer Matrix renormalization group (CTM) \cite{nishino1996,orus2009,corboz2011} algorithm
 to contract the network. The CTM algorithm approximates the entire lattice surrounding a bulk tensor by an environment, consisting of four corner tensors $C$ and four edge tensors $T$, as shown in Fig.~\ref{Fig:CTMsteps}(a-c). The accuracy is systematically controlled by the boundary bond dimension $\chi$ of these tensors. In the present work, due to the mirror symmetries of the bulk tensors, only a single $T$ tensor and two different $C$ tensors are needed. The latter two, which we label as $\{C_{\times}$,$C_{\diamond}\}$, differ by the orientation of the last absorbed bulk tensor, as defined in Fig.~\ref{Fig:CTMsteps}(e). Tensor $T$ is not mirror symmetric upon exchanging the boundary legs. Therefore, in order to keep track of its orientation, an oval in the shape of the last absorbed bulk tensor is added inside the depicted tensor in Fig.~\ref{Fig:CTMsteps}(e). All the boundary tensors are labelled by a superscript, which indicates the CTM iteration.

The initial boundary tensors $C_{\times/\diamond}^{(0)},T^{(0)}$ are constructed from the bulk $A$ and $B$ tensors by projecting the open boundary legs in the bra and ket layer onto a single state, e.g. onto the first element of each index, see Fig.~\ref{Fig:CTMsteps}(d). In each iteration of the CTM algorithm, two edge tensors and a bulk tensor are absorbed in the corner and a bulk tensor is absorbed in the edge tensor as shown in Fig.~\ref{Fig:CTMsteps}(f-g) thereby effectively growing the number of sites each boundary tensor represents. Both absorptions increase the boundary dimension from $\chi$ to $\chi \times D^2$, which is truncated back to the original boundary dimension $\chi$. The truncation is performed by two isometries $U_{\times}$, $U_{\diamond}$, which are obtained from a singular value decomposition (SVD) of the corresponding corner tensors, shown in Fig.~\ref{Fig:CTMsteps}(f)~\cite{nishino1996}. The algorithm is run for multiple iterations until the singular values $s$ of the corner matrix are converged. The total number of iterations needed we call $N$. 

\begin{figure}[tb]
\begin{center}
	\includegraphics[width=\columnwidth]{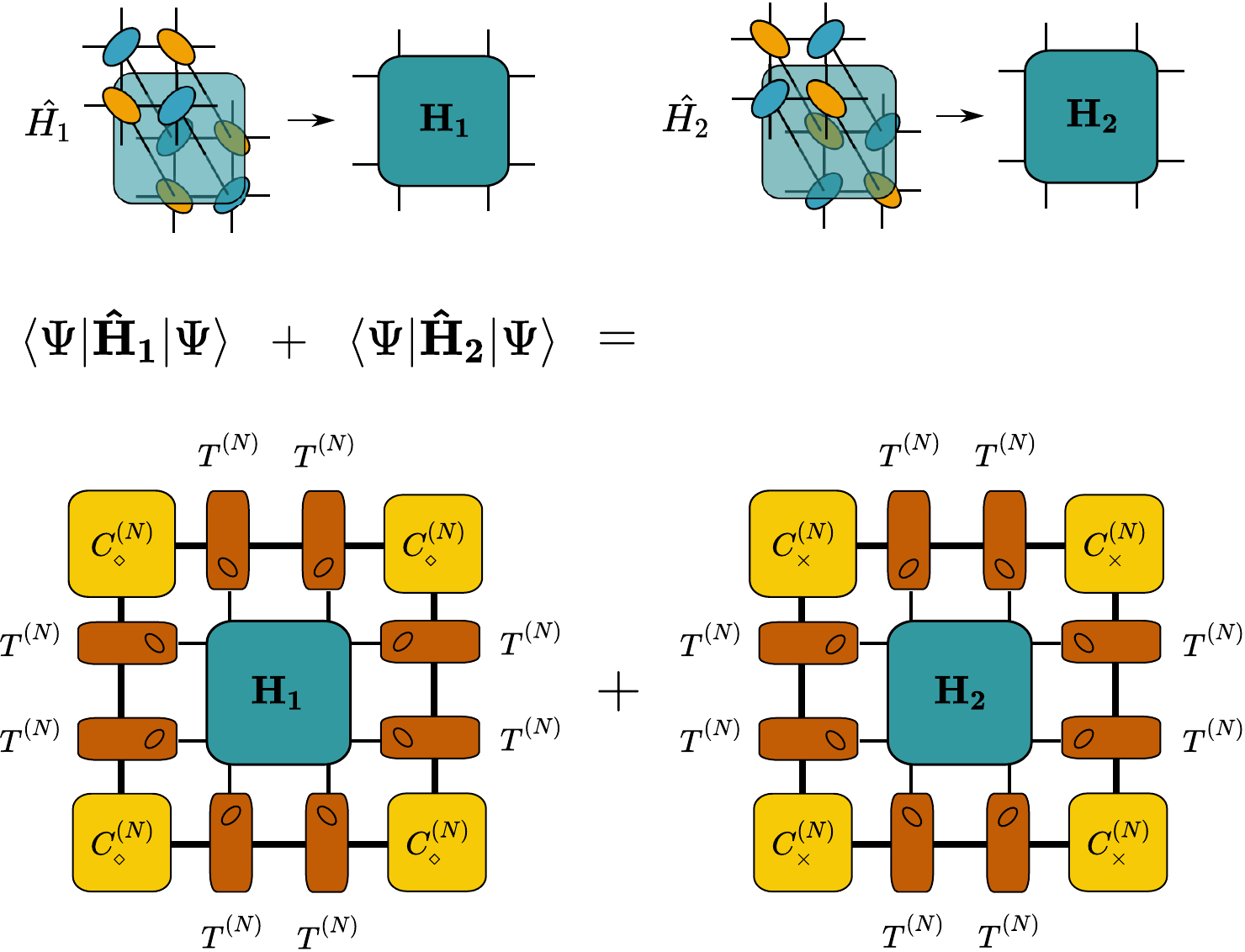}
\caption{The tensor network diagrams to evaluate the toric code Hamiltonian using the CTM environment tensors is shown. The plaquette contribution is labelled as $\hat{H_1}$ and the vertex contribution $\hat{H_2}$. }
\label{Fig:CTME}
\end{center}
\end{figure}

Using the converged CTM environment tensors, the energy of the toric code model can be evaluated as shown in Fig.~\ref{Fig:CTME}. We rewrite the Hamiltonian in Eq.~\ref{Eq:TC_Hamiltonianh} as $\hat{H} = \sum_v \hat{H_1} + \sum_p \hat{H_2}$, where $\hat{H_1}$ and $\hat{H_2}$ correspond to the vertex and plaquette terms, respectively,  where the on-site magnetic terms have been split evenly among both terms. Each of the two terms are then evaluated by making use of the corresponding CTM environment, see Fig.~\ref{Fig:CTME}. We note that the two tensors $\mathbf{H_1}$ and $\mathbf{H_2}$ defined in Fig.~\ref{Fig:CTME} are never explicitly constructed since this would be computationally inefficient.

\subsection{Optimization}
\label{sec:GradientOpt}
Recently, methods to optimize the iPEPS tensors based on an energy minimization have been introduced~\cite{corboz2016, vanderstraeten2016}. The goal is to find the optimal parameters in the tensors which minimize the variational energy, 
\begin{equation}
	\label{Eq:grad}
	E = \frac{\bra{\Psi} \hat{H} \ket{\Psi}}{\braket{\Psi}{\Psi}},
\end{equation}
in order to obtain the best approximation to the exact ground state of a Hamiltonian $\hat H$ for a given bond dimension $D$. In Refs.~\onlinecite{corboz2016, vanderstraeten2016} different approaches have been proposed to calculate the derivative of Eq.~\ref{Eq:grad} with respect to the tensors, which is then used either to perform a conjugate-gradient optimization~\cite{vanderstraeten2016} or to solve a generalized eigenvalue problem~\cite{corboz2016} to lower the energy in an iterative way. 
The derivative $\partial_{A^\dagger} \bra{\Psi} \hat{H} \ket{\Psi}$ can be written as a double infinite sum where  one sum goes over all Hamiltonian terms and the other sum over the locations of the "hole" created by taking the derivative with respect to $\partial_{A^\dagger}$ (note that the derivative of a tensor network with respect to a tensor $X$ is given by the network with tensor $X$ being removed). Due to translation invariance only the relative distances between the Hamiltonian terms and the holes matter, such that in practice only one of the two sums needs to be performed.  

\begin{figure*}[tb]
\includegraphics[width=\textwidth]{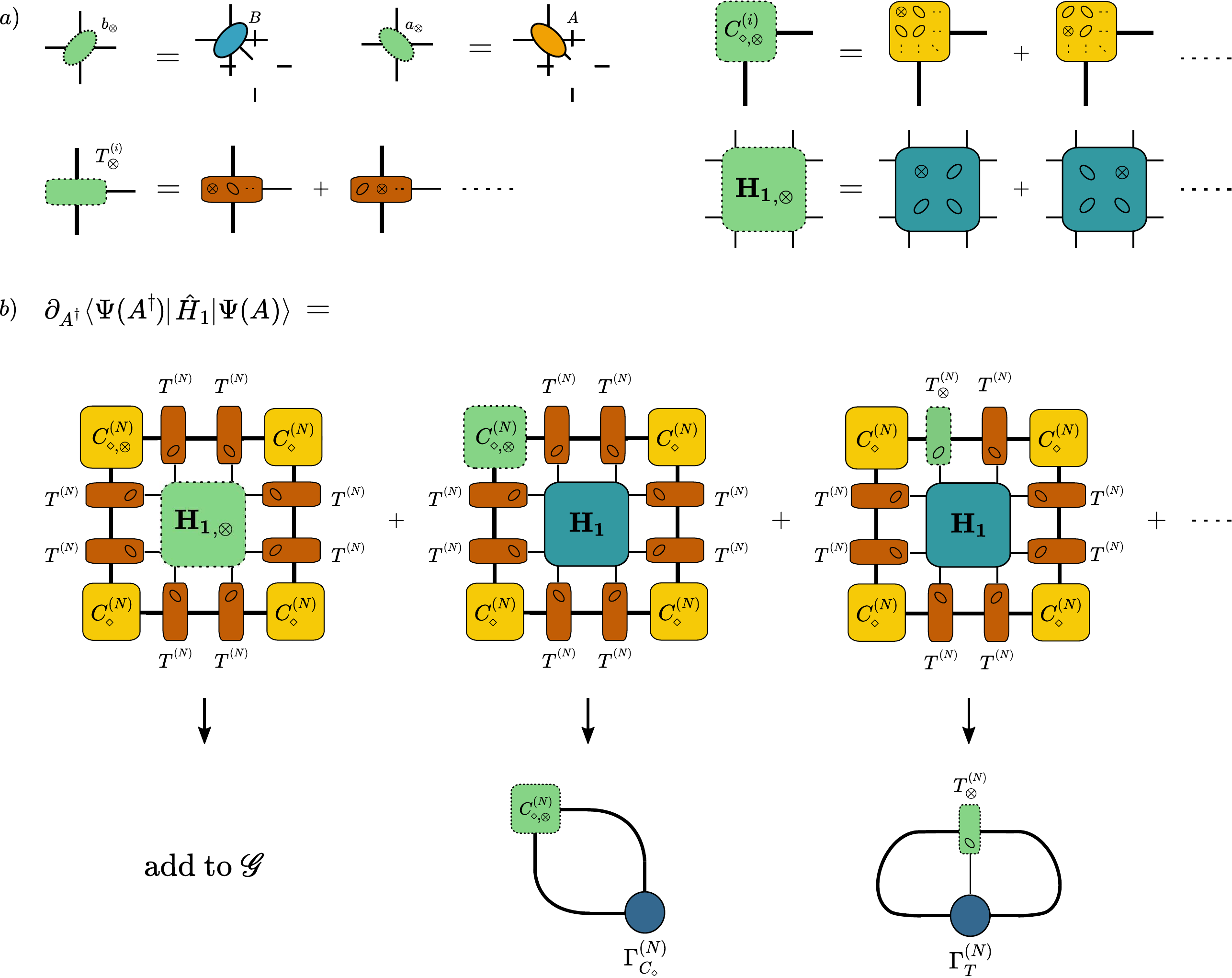}
\caption{Tensor network diagrams to represent the gradient of the energy with respect to tensor $A^{\dagger}$ (only the contributions from $\hat H_1$ are shown). (a) The derivative of the bulk and boundary tensors with respect to $\partial A^{\dagger}$ are represented by the green tensors, e.g.  $a_{\otimes} = \partial_{A^{\dagger}} a$. The  ${\otimes}$ symbol in the tensor shapes mark the different locations in the lattice at which the derivative is taken.
In $(b)$ the total derivative of $\langle \hat H_1 \rangle $ is represented in terms of a summation over the different tensors introduced in $(a)$. The first term can be directly evaluated and its contribution can be added to the total gradient $\mathscr{G}$. The other contributions are rewritten in terms of the environments $\Gamma_{C_\diamond}^{(N)},\Gamma_{T}^{(N)}$ and $C^{(N)}_{\diamond,\otimes}, T^{(N) }_{\otimes}$ which are evaluated recursively as shown in Fig.~\ref{Fig:OptSetup2}.}
\label{Fig:OptSetup1}
\end{figure*}

While in Refs.~\onlinecite{corboz2016, vanderstraeten2016} the summation was done over all Hamiltonian terms (with the hole kept fixed in the center), here we propose an alternative approach where we sum over all possible locations of the hole, with the Hamiltonian term(s) kept fixed in the center. This has the advantage that more complicated models, e.g. with longer ranged interactions and/or multi-site interactions (such as in the toric code model) can be treated more easily and computationally more efficiently, since no summation over these Hamiltonian terms is required, and the systematic summation over the hole position is model independent.

Following Ref.~\onlinecite{vanderstraeten2016}, we treat $A$ and $A^\dagger$ as independent and write the gradient as
\begin{align}
	\label{Eq:grad2}
	\partial_{A^\dagger} [\frac{\bra{\Psi} \hat{H} \ket{\Psi}}{\braket{\Psi}{\Psi}}] =
	\frac{\partial_{A^\dagger} \bra{\Psi} \hat{H} \ket{\Psi}}{\braket{\Psi}{\Psi}} - \frac{\bra{\Psi} \hat{H} \ket{\Psi}}{\braket{\Psi}{\Psi}^2} \partial_{A^\dagger} \braket{\Psi}{\Psi}.
\end{align}
The expression is simplified by shifting the Hamiltonian $\hat{H} \rightarrow  \hat{H} - \bra{\Psi} \hat{H} \ket{\Psi}$, such that only the first term in Eq.~\ref{Eq:grad2} remains. 

In the following we explain how to compute the gradient for the toric code model with two different 4-site Hamiltonian terms $\hat H_1$ and $\hat H_2$, but we stress  that the approach can also be easily applied to other type of Hamiltonians. The contributions to the gradient with respect to the $\hat H_1$ term are presented  in Fig.~\ref{Fig:OptSetup1}(b): the location of the $\hat H_1$ operator is fixed in the center and the sum is taken over all possible locations of the hole which is  located either in one of the corner tensors, one of  the edge tensors, or one of the four sites in the center. Each green tensor in Fig.~\ref{Fig:OptSetup1}(b) denotes a sum over all possible hole locations on the sites that the tensor is effectively representing, as defined in Fig.~\ref{Fig:OptSetup1}(a), and we label the corresponding tensors with an  $\otimes$ symbol.
We note that the explicit construction of the green tensors is computationally inefficient. Instead, we use a recursive scheme based on the CTM method to sum up all contributions as described in the following.
 
\begin{figure*}[tb]
\includegraphics[width=\textwidth]{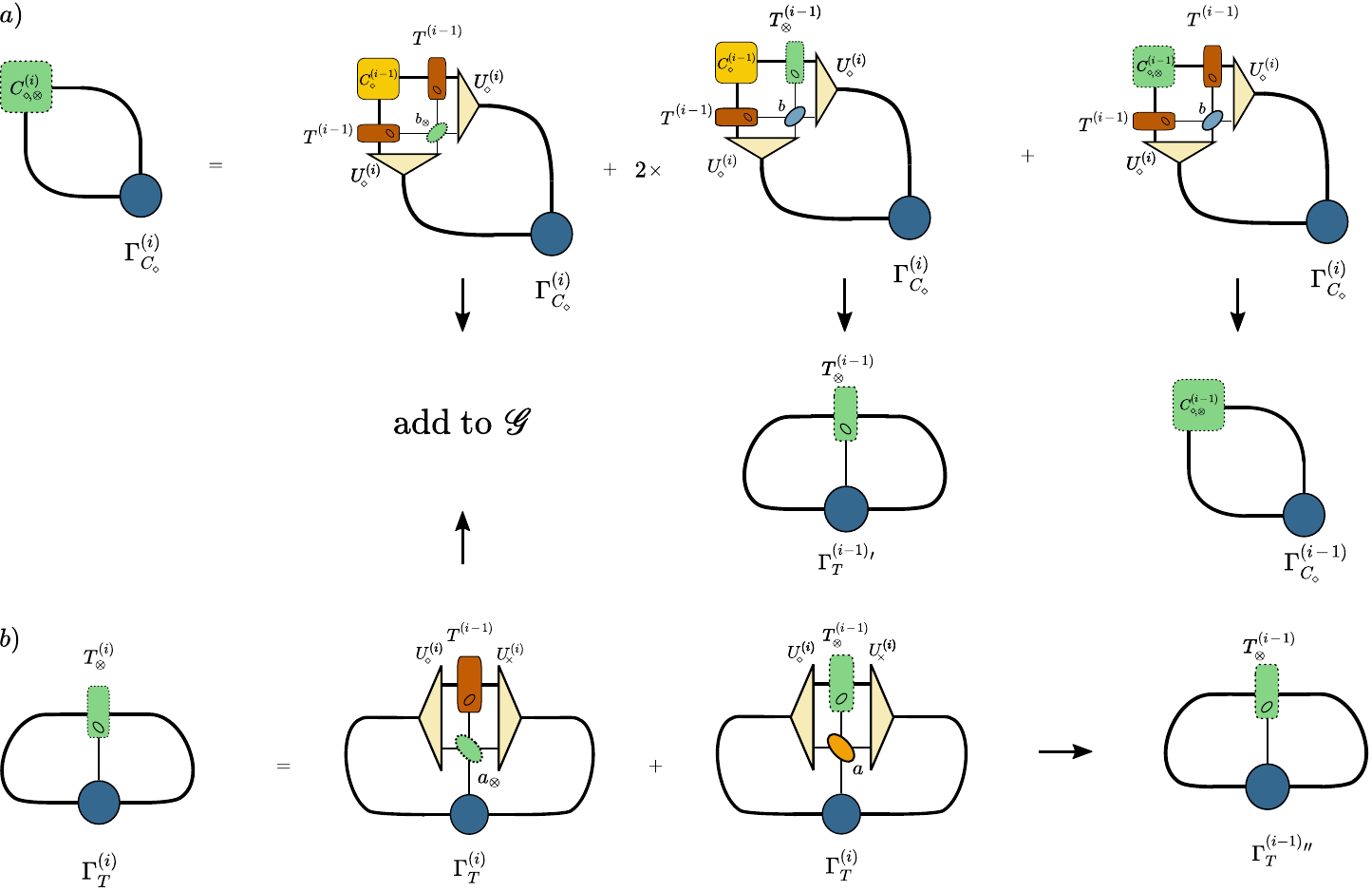}
\caption{Recursive procedure to sum up all contributions to the gradient (only the contributions from $\hat H_1$ are shown). (a) The corner gradient tensor $C^{(i)}_{\diamond,\otimes}$ is expanded in its four contributions. The first term can be directly evaluated and its contribution added to the total gradient $\mathscr{G}$. The second term, which is taken with a prefactor 2  due to the presence of two $T$ tensors which are equivalent by symmetry,  becomes a contribution $\Gamma_T^{(i-1)'}$ to the edge environment for the next iteration, and the last term generates the corner environment tensors for the next iteration, $\Gamma^{(i-1)}_{C_{\diamond}}$. 
Similarly, in $(b)$ the edge tensor $\Gamma^{(i)}_{T}$ is expanded into two terms, where the first term can be directly evaluated and added to the total gradient $\mathscr{G}$, and the second term yields a contribution $\Gamma_T^{(i-1)''}$ to the edge environment for the next iteration. (Note that in the second term the orientation of the boundary legs flips.) After each iteration the different $(i-1)$th edge environment contributions obtained in the corner and the edge update step (from  both $\hat H_1$ and $\hat H_2$) are added to obtain the  edge environment $\Gamma^{(i-1)}_{T}$ for the next iteration. }
\label{Fig:OptSetup2}
\end{figure*}

We start by applying the CTM method (Sec.~\ref{Sec:CTM}) until convergence is reached after $N$ iterations. All corner, edge, and isometry tensors at each CTM iteration are saved, and labelled by a superscript indicating the CTM iteration. Using the converged tensors  at the final iteration $N$, the first term in Fig.~\ref{Fig:OptSetup1}(b) can  directly  be evaluated and added to the total gradient $\mathscr{G}$. 
The idea is now to construct the remaining terms of the summation recursively in order to recover all contributions to the gradient in a systematic way. 
We  define the  environment tensors $\Gamma_{C_\diamond}^{(N)}$ and $\Gamma_{T}^{(N)}$ corresponding to  the tensor network surrounding the  $C^{(N)}_{\diamond, \otimes}$ and $T^{(N)}_{\otimes}$ tensors in Fig.~\ref{Fig:OptSetup1}(b), respectively. Starting from these environments at step $i=N$ we propagate backwards in the CTM steps and iteratively compute the respective environments at the $(i-1)$th step, and at the same time sum up all the contributions to the gradient, as described in Fig.~\ref{Fig:OptSetup2}. At each iteration, the $i$th corner (edge) gradient tensor contracted with its respective environment is rewritten in terms of the corner (edge) gradient tensor of the $(i-1)$th step plus a one-site gradient contribution  coming from a bulk tensor which is added to the total gradient $\mathscr{G}$. A similar recursion is done (simultaneously) for the other Hamiltonian term $\hat H_2$, involving the environment tensor $\Gamma_{C_\times}^{(i)}$ and corner gradient tensor $C^{(i)}_{\times, \otimes}$. The procedure is repeated iteratively until either all the $N$ CTM steps have been recovered, or until the total gradient $\mathscr{G}$ is converged within a certain tolerance.

The total gradient can then be used in combination with a gradient-based minimization algorithm to optimize the tensors. In this work we used the  Broyden-Fletcher-Goldfarb-Shanno (BFGS) quasi-newton method~\cite{broyden1970,fletcher1970,goldfarb1970,shanno1970}. The method is converged when either the norm of the gradient becomes smaller than a certain tolerance, or when it is no longer possible to find a suitable step-size which would decrease the energy. In practice, in order to prevent convergence to a local minimum, we run the optimization starting from several random initial tensors and keep the state with the lowest variational energy.

This method has several advantages over the optimization schemes based on the summation of Hamiltonian terms~\cite{corboz2016, vanderstraeten2016}. First, the iterative part of the algorithm where the environments are updated scales in the same way as the normal CTM method, $O(\chi^3 D^4) + O(\chi^2 D^6)$. The dominant scaling of the algorithm lies in the calculation of the initial environments $\Gamma_{C_\times/C_\diamond}^{(N)}, \Gamma_{T}^{(N)}$, which, due to the four body Hamiltonian, scales as $O(\chi^3 D^6)$, but it only needs to be done once per gradient computation. This is in contrast to previous approaches~\cite{corboz2016, vanderstraeten2016} involving the summation of  four-body contributions at each CTM iteration which is computationally less efficient. Second, the current scheme can be very easily extended to other types of Hamiltonians, including longer ranged and multi-site interactions, by initializing the initial environments accordingly (i.e. without performing a summation of more complicated Hamiltonian terms). 

We note that  recently, an alternative optimization method based on automatic differentiation (AD) has been introduced~\cite{liao2019}. This method performs a similar back-propagation of environments as in our CTM approach in an automatized fashion. One difference is that the AD approach also includes the  gradient contribution with respect to the SVD step in the CTM method in contrast to our approach. It would be interesting to compare the performance of the two approaches in  detail, which we leave for future work.


\section{Detection of a topologically ordered phase with iPEPS}
\label{chap:Framework}

\subsection{Virtual symmetry} 
\label{Sec:TopIpeps}
\begin{figure}[tb]
\begin{center}
\includegraphics[width=\columnwidth]{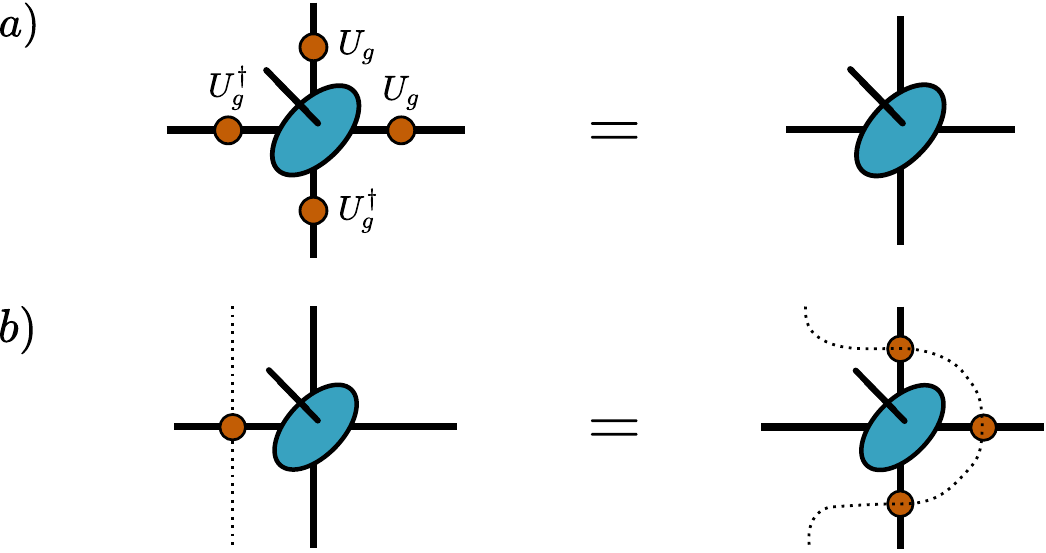}
\caption{(a) A tensor with a virtual symmetry, i.e. which is  invariant under the simultaneous action of $U_g$ on the virtual level.  This naturally leads to the "pulling-through" condition shown in b). }
\label{Fig:PullingThrough}
\end{center}
\end{figure}

The advantage of using  iPEPS to study (non-chiral) topologically ordered phases is that there exist a powerful framework on how these phases can be represented with this ansatz~\cite{schuch2010}. The distinguishing properties of a topologically ordered phase occur at a global level, however it has been shown that these properties can be translated to necessary symmetry requirements on the virtual indices of the  iPEPS tensors, namely 
that the tensors are invariant under the action of a certain symmetry group $G$ on the virtual indices of the tensors, which is called a virtual symmetry. More specifically, 
the tensor remains invariant under the simultaneous action of $U_g$ on all the virtual indices, where $U_g$ is a unitary representation of an element $g$ of the group $G$, as shown as shown in Fig.~\ref{Fig:PullingThrough}(a). This idea has also been generalized to symmetries represented by matrix-product operators~\cite{Buerschaper2014,sahinoglu2014} obeying a similar pulling-through condition as shown in Fig.~\ref{Fig:PullingThrough}(b), i.e. pulling an MPO (or in our case a product of operators on a string) through tensor leaves the tensor invariant.

Here we will  focus on the \ZZ topologically ordered phase of the toric code model where the tensors are invariant under the simultaneous action of a unitary representation of the \ZZ symmetry group on the virtual legs. In the $D=2$ case, we use the standard representation $U_g \in \{\mathbf{I},\sigma^z\}$, with a straightforward generalization to larger $D$. To each state in the virtual space we can assign a parity label even (e) or odd (o) and the symmetry condition with $U_g=\sigma^z$ implies that elements with a total odd parity in a tensor are vanishing, leading to a block structure of the tensors.  

This block structure can be easily seen in the exact representation of the TC ground state,  by identifying the first and second element of the virtual indices with the even and odd sectors, respectively. It naturally arises from the constraint of having closed loops in  the ground state, i.e. whenever a loop (odd parity) enters a tensor, it has to exit it again, and all tensor elements corresponding to an odd total parity are zero.

\subsection{Challenges in practical simulations} 
We have seen in the previous section that the iPEPS tensors representing the exact TC ground state exhibit a  virtual \ZZ symmetry, i.e. a block structure where half of the tensor elements are zero due to this symmetry. However, in practice, when performing an optimization of the tensors for the toric code Hamiltonian starting from random initial tensors, the resulting tensors will in general not be \ZZ symmetric for two reasons. First, the symmetry may not be apparent due to the gauge freedom in the tensor network, i.e. between each bond two $D \times D$ size matrices $q$ and  $q^{-1}$ can be inserted,  acting as a basis transformation on the virtual level, which does not  change the state. After the optimization the tensors are in an arbitrary gauge, i.e. not necessarily in the basis in which they  are \ZZ symmetric. 

The second reason is that the optimization may fail to yield perfectly symmetric tensors due to round-off and truncation errors and non-perfect convergence. Even if these errors are small they may be problematic since it has been shown that already small errors which violate the \ZZ virtual symmetry lead to a loss of topological order~\cite{chen2010,shukla2018}. One way to overcome this problem is to enforce the virtual symmetry during the optimization~\cite{he2014a, shukla2018}, which however requires knowledge of the correct virtual symmetry beforehand. (If it is not known, one would need to run separate simulations, testing different virtual symmetries, which is  not efficient since the optimization is computationally the most expensive part).

In order to overcome these issues, in the following we present a scheme to recover the virtual symmetry starting from an unconstrained optimization. This allows us to perform an unbiased iPEPS optimization, i.e. without a priori imposing a virtual symmetry, and we will show that it is possible to correctly identify the topological order. 


\subsection{Restoring the virtual symmetry}
\label{Chap:RestorSym}
\begin{figure}[tb]
\begin{center}
\includegraphics[width=\columnwidth]{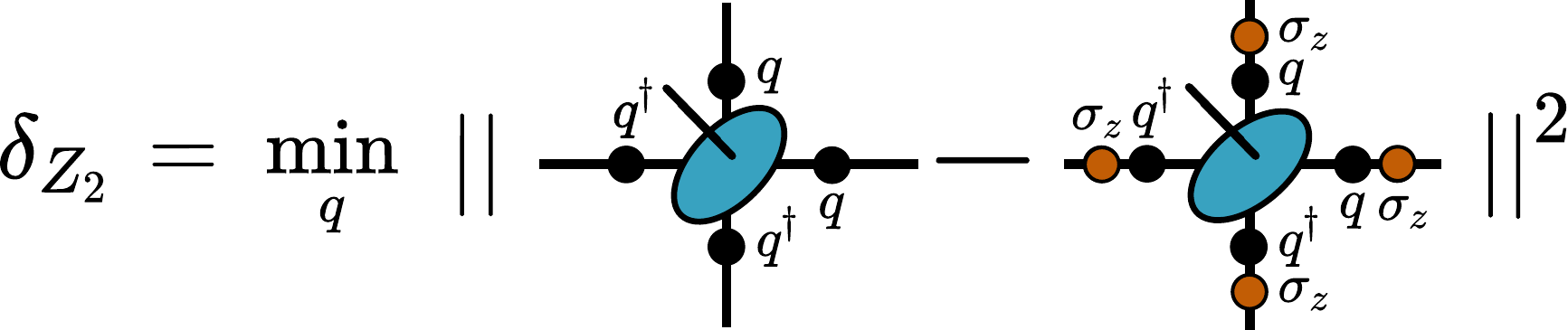}
\caption{The PEPS tensor is brought into an approximate \ZZ symmetric form by performing a gauge change using a unitary matrix $q$ which minimizes $\delta_{Z_2}$.
}
\label{Fig:MinimizeGauge}
\end{center}
\end{figure}
In this section we present a scheme to restore the \ZZ virtual symmetry of the tensors obtained from an unconstrained optimization (i.e. where we do not impose the virtual symmetry during the optimization).
We start by fixing the gauge freedom $q q^{-1}$ between all the tensors such that in the final basis the tensor is closest to a \ZZ symmetric tensor. Due to the mirror symmetries on our $A$ and $B$ tensors, $q$ reduces to a $D\times D$ unitary matrix. To fix the gauge we have to pick a basis in which we define the \ZZ symmetry. In practice, for a bond dimension~$D$, we choose a basis where we associate the first $D/2$ entries of a leg with the even sector, and the rest with the odd sector. When $D$ is odd, we round the  dimension up (down) to an integer for the even (odd) sector. The initial $q$ is taken as a random unitary matrix. Then, the optimal $q$ is found by minimizing the norm difference $\delta_{Z_2}$ as shown in Fig.~\ref{Fig:MinimizeGauge}, which measures the deviation of the tensor from a \ZZ symmetric one. If $\delta_{Z_2}$ is zero, it implies that the tensor fulfills the condition in Fig.~\ref{Fig:PullingThrough}(a), i.e. that it is perfectly \ZZ symmetric. The optimization of $q$ is done by a quasi-Newton minimization. Once convergence is reached, we transform the tensor using the $q$ matrices, yielding a new tensor which is approximately \ZZ symmetric.

After fixing the gauge, the tensor $A$ (and $B$) exhibits a block structure, in which the elements lying outside the allowed \ZZ symmetry blocks are small. By setting these small elements to zero, the tensor becomes fully \ZZ symmetric, leading to a state with a robust topological order (if we consider a state within the topologically ordered phase). This procedure should not alter the state in any significant way, e.g. it should only lead to a very small difference in energy. We will show in the results section Sec.~\ref{chap:Results} that this is indeed the case. 


\subsection{Computing the TEE}
\label{Chap:ComputingTEE}
We obtain the TEE from the second R\'{e}nyi entropy~\cite{flammia2009}, $S_2(\rho_L) = - \log [\mathrm{Tr}(\rho_L^2)]$, between two halves $L, R$ of an infinite cylinder which can be efficiently computed with iPEPS based on ideas from Ref.~\onlinecite{cirac2011}. It has been shown that there exists an exact mapping between the physical degrees of freedom of a region and the virtual degrees of freedom connecting to this region. Specifically, as shown in Ref.~\onlinecite{cirac2011}, the reduced density matrix $\rho_L$ can be represented as
\begin{equation}
\label{eq:rhol}
	\rho_L = U \sqrt{\sigma_L^{T}} \sigma_R \sqrt{\sigma_L^{T}} U^{\dagger}
\end{equation}
where $\sigma_L$ and $\sigma_R$ are the left and right reduced density operators defined in the virtual space along the cut, respectively. That is, $\sigma_L$ ($\sigma_R$) is obtained by contracting the double-layer tensor network in the region $L$ ($R$),  keeping the virtual indices at the boundary open. $U$ is an isometry defining the mapping between the physical and virtual space. With Eq.~\ref{eq:rhol} the second R\'{e}nyi entropy can be written as 
\begin{equation}	
	\label{eq:renyiEE}
	S_{2}(\rho_L) = - \log[ \mathrm{Tr} (\sigma_L^{T} \sigma_R \sigma_L^{T} \sigma_R) / N^2],
\end{equation}
where we have introduced a normalization factor $N = \mathrm{Tr} [\rho_L] = \mathrm{Tr} [\sigma_L \sigma_R^{T}]$ for the case that $\rho_L$ is not normalized.

\begin{figure}[tb]
\begin{center}
\includegraphics[width=\columnwidth]{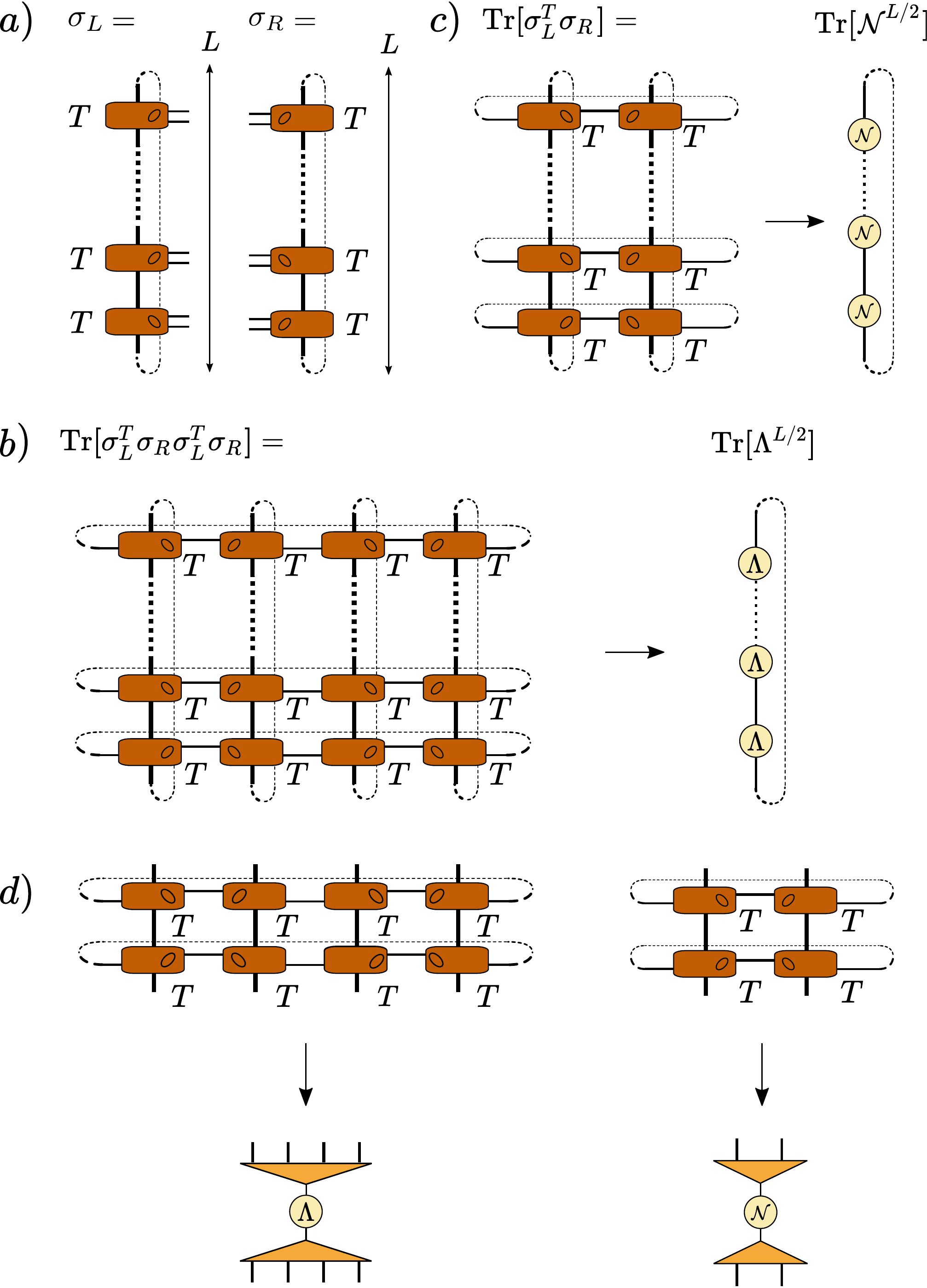}
\caption{Diagrams for the computation of the second R\'{e}nyi entropy between two halves of an infinite cylinder of width~$L$. (a) Representation of $\sigma_L$ and $\sigma_R$ of the left / right half of the infinite cylinder in terms of the edge tensors $T$ obtained from the CTM method. (b)-(c) Tensor networks representing the traces in Eq.~\ref{eq:renyiEE}. Instead of contracting a large network we diagonalize the double row of $T$ tensors shown in (d) to obtain the eigenvalue matrices $\Lambda$ and $\mathcal{N}$, so that the traces in (b) can be computed as $\mathrm{Tr} [\Lambda^{L/2}]$ and $\mathrm{Tr} [\mathcal{N}^{L/2}]$, respectively.
}
\label{Fig:CylinderCalculationNetworks}
\end{center}
\end{figure}

Using the CTM algorithm, $\sigma_L$ (and $\sigma_R$) of an infinite cylinder of circumference $L$ can  simply be represented by a periodic chain of edge tensors $T$ as shown in Fig.~\ref{Fig:CylinderCalculationNetworks}(a), since each edge tensor represents an infinite row of bulk tensors contracted with its complex conjugate.
The trace in Eq.~\ref{eq:renyiEE} can then be obtained by contracting the tensor network shown in Fig.~\ref{Fig:CylinderCalculationNetworks}(b), made of a $4\times L$ periodic network of $T$ tensors (taking into account the orientation of the $T$ tensors), and the normalization factor is represented in Fig.~\ref{Fig:CylinderCalculationNetworks}(c).  A similar approach was also used in Refs.~\onlinecite{orus2014b,jahromi2018}. For large cylinders, it is beneficial to diagonalize the matrices represented by the two rows of edge tensors shown in Fig.~\ref{Fig:CylinderCalculationNetworks}(d), and then obtain the second R\'{e}nyi entropy as $S_{2}(L) = - \log[\mathrm{Tr}(\Lambda^{L/2}) / [\mathrm{Tr}(\mathcal{N}^{L/2})]^2]$, where $\Lambda$ and $\mathcal{N}$ are the corresponding diagonal matrices.

The TEE entanglement entropy can then be determined  from the intersection of a linear fit to $S_{2}(L)$ for sufficiently large L with the y-axis at $L=0$, see Fig.~\ref{Fig:LargeCylinder}  in Sec.~\ref{chap:Results} for an example. $S_{2}(L)$ is computed from one of the MES, which we  obtain as explained in the following section.


\subsection{Calculation of the modular S and U matrices with CTM}
\label{Chap:Calculation_CTM_matrices}
In Ref.~\onlinecite{he2014a} an approach based on  TRG was introduced for the calculation of wave function overlaps to determine the modular S- and U-matrices. The idea is to preserve the virtual symmetry in the TRG coarse-graining process in both the bra- and ket-layer, such that the resulting coarse-grained tensor $\mathbb{T}$ representing the infinite 2D system on a torus exhibits the same virtual symmetry. From this tensor all the MES can be obtained by acting with operators on the virtual legs, and overlaps between them can be efficiently computed. 

Here we introduce an alternative scheme based on the CTM, which is computationally more efficient than TRG. For simplicity we discuss it here for the case of a \ZZ topological order, and we consider bulk tensors with a \ZZ virtual symmetry (e.g. after applying the symmetrization procedure described in Sec.~\ref{Chap:RestorSym}). The main idea is to keep track of the parity sectors on the open boundaries during the CTM iterations, such that the different ground states can be individually selected on the bra and ket level, similarly as in the TRG approach~\cite{he2014a}.

\begin{figure}[tb]
\begin{center}
\includegraphics[width=\columnwidth]{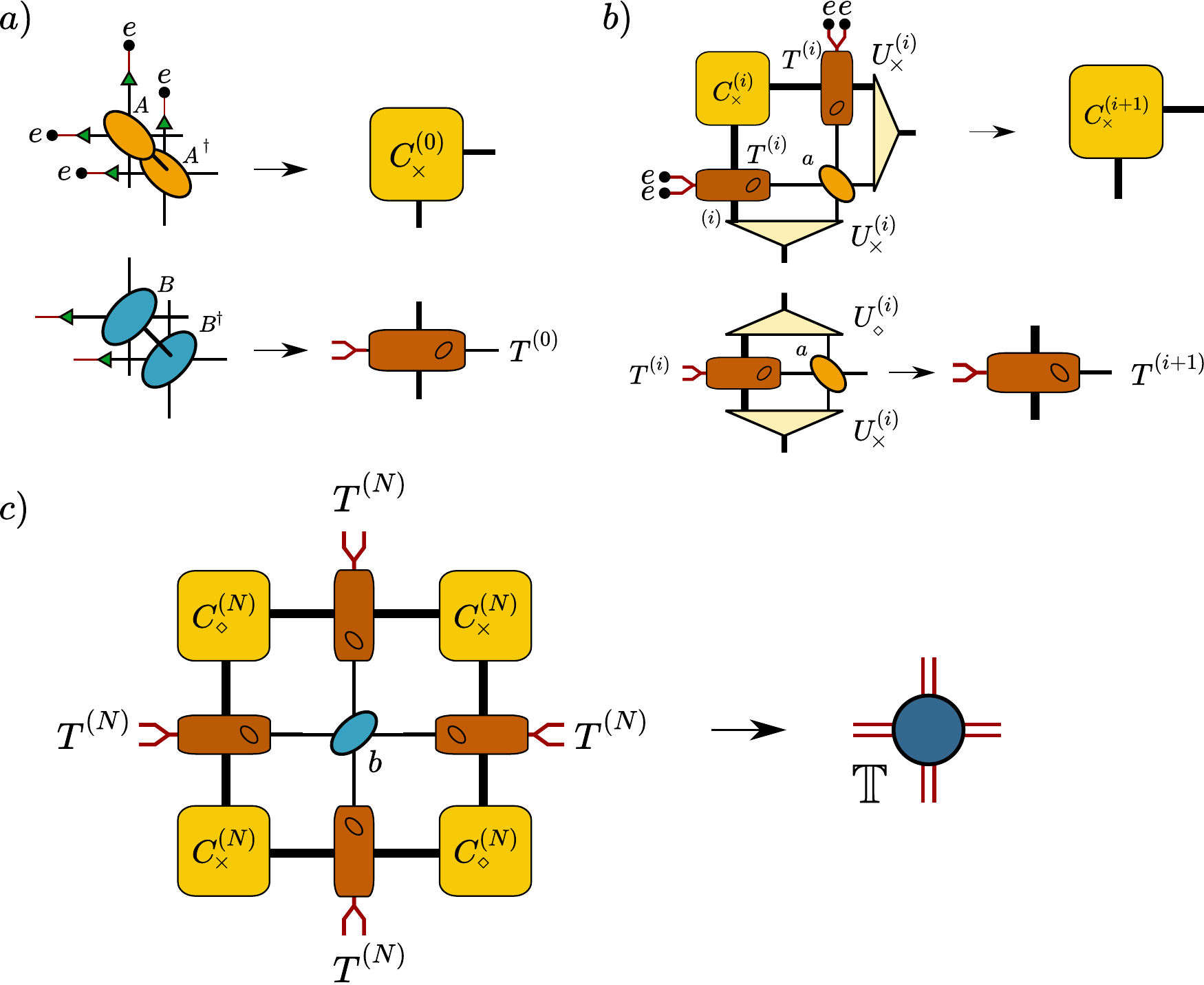}
\caption{Variant of the CTM algorithm where the parity on the bra- and ket- level on each boundary can be controlled. The red lines are of dimension 2, with one state in the even (e) and odd (o) sector, respectively, and the green triangles denote a projection onto this space. A black dot corresponds to a projection onto the even parity sector. (a) Initialization of the boundary tensors, where the boundaries of the corner tensor is projected onto the even-even sector, and the red parity legs of the $T$ tensor are kept open. (b) Coarse-graining step, in which the corner tensors are kept in the even-even boundary sector, and the red parity legs of the $T$ remain open. (c) Contraction of the network on the left yields the double-layer tensor $\mathbb{T}$ representing an infinite plane (or infinite torus when connecting the legs in a periodic way), where the parity in each layer on each boundary can be controlled via the red legs.}
\label{Fig:ParityCTM}
\end{center}
\end{figure}

The following adaptations are made to the standard CTM algorithm (Sec.~\ref{Sec:CTM}): first, instead  of initializing the open boundary of the edge tensor $T$ by a projection onto a single state we keep two states, one with even and one with odd parity, on both the bra and ket level, as shown in Fig.~\ref{Fig:ParityCTM}(b). These extra legs are kept open  in each CTM renormalization step (Fig.~\ref{Fig:ParityCTM}). Second, we initialize the corner tensor to have an even-even parity on the two open boundaries, and we keep it in this sector in each  renormalization step by projecting all of the open parity legs of the absorbed $T$ tensors onto the even-even sector (Fig.~\ref{Fig:ParityCTM}(b)). The isometries to perform the renormalization are found in a similar way as in the standard approach. 

Once the CTM has converged we construct the network in Fig.~\ref{Fig:ParityCTM}(d) representing the infinite 2D system. Since the parity on the boundary of the corner tensors is fixed to the even-even sector, the total parity on the boundaries of the bra and ket level can be fully controlled by the extra legs of the edge tensors. Contracting the network yields a  double-layer tensor $\mathbb{T}^{ijkl}_{i'j'k'l'}$ with a \ZZ $\times$ \ZZ symmetry similar to the one obtained with TRG in  Ref.~\onlinecite{he2014a}. Further, a torus geometry can be mimicked by connecting the horizontal and vertical legs in a periodic way. Specific ground states can be obtained by projecting the boundary onto the desired sectors in horizontal and vertical direction. We normalize the tensor $\mathbb{T}$ in each sector such that taking the trace yields an equal superposition of  the four ground states, $( \ket{\Psi_{ee}}+\ket{\Psi_{eo}}+\ket{\Psi_{oe}}+\ket{\Psi_{oo}})$, in the bra and ket layer.  

From this all MES can be constructed, by using Eq.~\ref{Eq:MESs}, or (equivalently) by finding the eigenstates of  the loop operators $W^{(z)}(C)$ and $W^{(x)}(C)$ which on the level of the $\mathbb{T}$ tensor are simply given  by a single $\sigma^z$ and $\sigma^x$ acting on a virtual leg of the $\mathbb{T}$ tensor. From the MES, the S and U matrices can then be computed, similarly as done in Ref.~\onlinecite{he2014a}.


\section{Results}
\label{chap:Results}
\subsection{Toric code model without magnetic field}

We start by testing our approach for the  toric code model without magnetic field ($h_x = h_z = 0$, with $J=1/2$). The optimization is done for $D=2$ and $\chi=40$. Since there exists an exact $D=2$ representation, we expect to be able to reproduce the  ground state accurately. The CTM contraction is terminated when the difference in the spectrum of the subsequent corner tensors $ ||s^{(i+1)} - s^{(i)} || < 10^{-12}$. The minimization is done using the gradient-based energy minimization approach described in Sec.~\ref{sec:GradientOpt}. 
 At each iteration,  the calculation of the gradient is completed either when all the CTM steps are recovered or when the total gradient has converged, $||\mathscr{G}^{(i)} - \mathscr{G}^{(i-1)}|| <  10^{-12}$.
The optimization is terminated when either the difference in energy between two iterations has become smaller than $10^{-14}$, or when it is no longer possible to find a new direction which lowers the energy. We run several independent optimizations with different random initial states and take the state with lowest variational energy. The optimization scheme yields an iPEPS with only a very small difference in energy compared to the exact result,  $|E - E_{exact}| = 2.4\e{-9}$.

Next, we apply the schemes presented in Sec.~\ref{chap:Framework} to see whether we can detect the \ZZ topological order. We first determine the gauge in which the tensors are closest to a \ZZ symmetric form using the method described in Sec.~\ref{Chap:RestorSym}. We find that, after performing the gauge transformation, the tensor elements which lie outside the \ZZ virtual symmetry blocks are small. We quantify the deviation from a \ZZ symmetric tensor by $\delta_{Z_2}$ (cf. Fig~\ref{Fig:MinimizeGauge}) divided by the norm of the tensor. Here we obtain  $\delta_{Z_2}/||A||^2 = 1.6\e{-5}$, demonstrating that the tensors are very close to a \ZZ symmetric form, but they are not perfectly symmetric. After performing the symmetrization, i.e. setting the small elements which violate the \ZZ symmetry to zero (cf. Sec.~\ref{Chap:RestorSym}), we obtain a state which has essentially the same energy, $|E - E_{Z_2}| = 4.1\e{-10}$, showing that the symmetrization step has only a minor effect on the ground state energy.

\begin{figure}[tb]
\begin{center}
\includegraphics[width=\columnwidth]{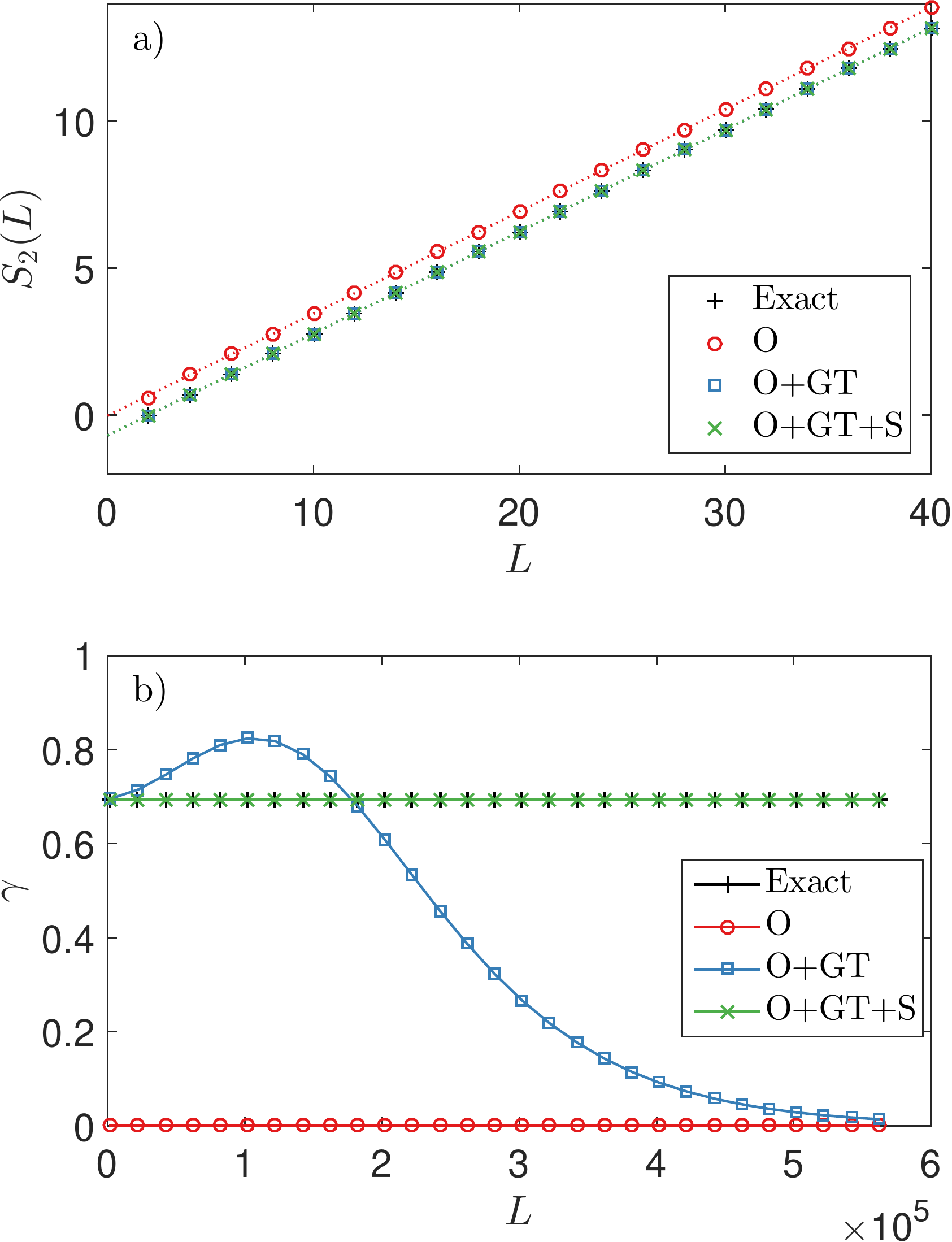}
\caption{(a) Second R\'enyi entropy of the toric code model as a function of width $L$ of an infinite cylinder, obtained from the optimized tensors ($O$), combined with a gauge transformation to bring the tensors into an approximate \ZZ symmetric form ($O+GT$), and symmetrized to recover the \ZZ virtual symmetry ($O+GT+S$). The TEE $\gamma$ is obtained from the intersection of a linear fit (dashed lines) with the y-axis. At short distances the correct value of TEE is obtained after performing the gauge transformation even without symmetrizing the tensors. (b) The TEE $\gamma$ obtained  from  linear fits to the data between $L$ and $L+100$ on large cylinders. At large distances, only the symmetrized tensors yield the correct value for $\gamma$. }
\label{Fig:LargeCylinder}
\end{center}
\end{figure}

In Fig.~\ref{Fig:LargeCylinder}(a) we present results for the second R\'enyi entropy as a function of the width $L$ of an infinite cylinder. To compute the TEE $\gamma$ we determine the intersection of a linear fit to the data with the y-axis. Using the tensors directly obtained from the optimization (labelled $O$), the TEE is not correctly reproduced, which is expected since the iPEPS is in an arbitrary gauge, i.e. it does not represent a MES. 
After performing the gauge transformation (labelled $O+GT$) we obtain the correct TEE with a deviation of only $1.7\times10^{-7}$
if we take the linear fit up to $L=40$. However, when increasing $L$  the result for the TEE starts to deviate from the exact result, see Fig.~\ref{Fig:LargeCylinder}(b), because they are not perfectly \ZZ symmetric. This is consistent with previous observations~\cite{chen2010} where a breakdown of the topological order was found by manually adding small perturbations to the tensors which violate the \ZZ virtual symmetry. In contrast, after symmetrization (labelled $O+GT+S$) the TEE is correctly reproduced even on very large cylinders~\footnote{we note that due to round off and truncation errors during the contraction in the CTM approach, the \ZZ symmetry could become broken in the environment which we observe here only when going to much larger cylinders. This problem could be circumvented by enforcing the \ZZ symmetry also in the environment tensors (not during the optimization, but only after the symmetrizing the iPEPS tensors),  as done in Ref.~\onlinecite{he2014a}.}.

Another difference between the $O+GT$  and the $O+GT+S$ simulations can be identified by comparing the eigenvalue spectrum of the transfer matrix, i.e. the eigenvalues $\mathcal{N}$ shown in  Fig.~\ref{Fig:CylinderCalculationNetworks}. For a \ZZ topologically ordered state we expect a twofold degeneracy~\cite{schuch2013}. While in the symmetrized case the degeneracy is accurate up to machine precision, without symmetrization the difference between the two leading eigenvalues is $2.8\times10^{-5}$, i.e. very close but not a perfect degeneracy, which causes the loss of topological order at long distances. 

\begin{table}
\begin{tabular}{c | c}
& $h_x = h_z = 0$ \\
\hline
 $E$ & $-0.9999999976 $\\
 \hline
 $|E - E_{Z_2}|$  &  $4.1\e{-10}$\\
 \hline
 $ \gamma - \log{2}$  & $1.1\e{-13}$\\
\hline
$\delta_{Z_2} / ||A||^2$ & $1.6\e{-05}$ \\
\hline
$S$ & $ \dfrac{1}{2} \begin{bmatrix}
    1 &   \phantom{-}1 	&   \phantom{-}1 	&  \phantom{-}1 \\
    1 &   \phantom{-}1 	&  -1 				&  -1 \\
    1 &  -1 			&  \phantom{-}1 				&  -1 \\
    1 &  -1 			&  -1 				&   \phantom{-}1 \\
\end{bmatrix}$\\
\hline
$U$ & $ \phantom{\dfrac{1}{2}} \begin{bmatrix}
	1  &  \phantom{-}0  & \phantom{-}0 &  \phantom{-}0 \\
    0  &  \phantom{-}1  & \phantom{-}0 &  \phantom{-}0 \\
    0  &  \phantom{-}0  & \phantom{-}1 &  \phantom{-}0 \\
    0  &  \phantom{-}0  & \phantom{-}0 &  -1 \\
    \end{bmatrix}$\\
\end{tabular}
\caption{Summary of the results obtained for the toric code model (without magnetic field) for $D=2$ (see main text for discussion).}
\label{Tab:h0}
\end{table}

Last, we calculate the modular $S$ and $U$ matrices based on the procedure explained in Sec.~\ref{Chap:Calculation_CTM_matrices} using the symmetrized tensors. We find values, presented in Tab.~\ref{Tab:h0}, which are in perfect agreement with the exact results (see e.g. Ref~\cite{zhang2012}), i.e. the anyonic particles $\mathbb{1}, e, m$ have bosonic self-statistics, whereas the $\epsilon$ particle has fermionic self-statistics, and braiding an $e$ particle with an $m$  particle (or with an $\epsilon=em$ particle) yields a phase of $\pi$ (i.e. they are mutually semions).

\subsection{Toric code model in a magnetic field}
Having established that the correct features of the topological ordered phase can be extracted in the unperturbed toric code model, we now test our approach for more challenging cases which are no longer exactly solvable. We first consider two  examples with a magnetic field applied in the z direction for  values $h_z = \{0.1,0.18\}$ which lie inside and outside the topologically ordered phase, respectively. (The location of the critical point is $h_z=0.164237(2)$~\cite{wu2012}).
 The simulations and analysis are done in a similar way as in the previous case, with the results summarized in Tab.~\ref{Tab:ComparissonD2} and Tab.~\ref{Tab:ComparissonOnlyHzD3}, for $D=2$ and $D=3$, respectively.

\begin{table}[tb]
\begin{tabular} {c | c  c  c}
 &\hsp $h_z = 0.10$ \hsp & \hsp$h_z = 0.18$ \hsp\\
\hline
 $E$  &   \hsp $ -1.01041 $  \hsp &\hsp  $ -1.04216$ \hsp \\
 \hline
 $|E - E_{Z_2}|$   & \hsp $ 5.7\e{-11} $ \hsp &  \hsp $ 6.8\e{-06} $ \hsp\\
  \hline
 $ \gamma - \log{2}$ & \hsp $7.0\e{-14}  $  \hsp& $ \hsp 0.6931$ \\
\hline
$ \delta_{Z_2} /  ||A||^2 $ \hsp &\hsp  $  7.5\e{-11} $ \hsp & \hsp $ 4.6\e{-3} $ \\
%
\end{tabular}
\caption{$D=2$, $\chi=40$ results for the toric code model with a magnetic field in z-direction. The data in the left and right column corresponds to a state in and outside of the topologically ordered phase, respectively. In the topologically ordered phase the same $S$ and $U$ matrices are obtained as in Tab.~\ref{Tab:h0}, with an accuracy which is close to machine precision (cf. Fig.~\ref{Fig:ErrorSUmatrices}).
}
\label{Tab:ComparissonD2}
\end{table}

\begin{table}[tb]
\begin{tabular} {c | c  c }
 &   \hsp $h_z = 0.10$   \hsp &   \hsp $h_z = 0.18$   \hsp\\
\hline
$E$   \hsp &   \hsp $-1.01042$  \hsp &   \hsp$ -1.04220$\\
\hline
 $|E - E_{Z_2}|$     \hsp&   \hsp$ 1.7\e{-06}$   \hsp&  \hsp $ 1.5\e{-5} $\\
  \hline
 $\gamma - \log{2} $   \hsp&  \hsp $ 7.3\e{-15} $   \hsp&  \hsp $ 0.6931$ \\
 \hline 
 $\delta_{Z_2} / ||A||^2$   \hsp&  \hsp  $6.3\e{-2} $   \hsp&  \hsp $4.8\e{-2}$  \\ 
\end{tabular}
\caption{Same as in Tab.~\ref{Tab:ComparissonD2}, here for $D=3$, $\chi=60$.
}
\label{Tab:ComparissonOnlyHzD3}
\end{table}

\begin{figure}[tb]
\begin{center}
\includegraphics[width=1\columnwidth]{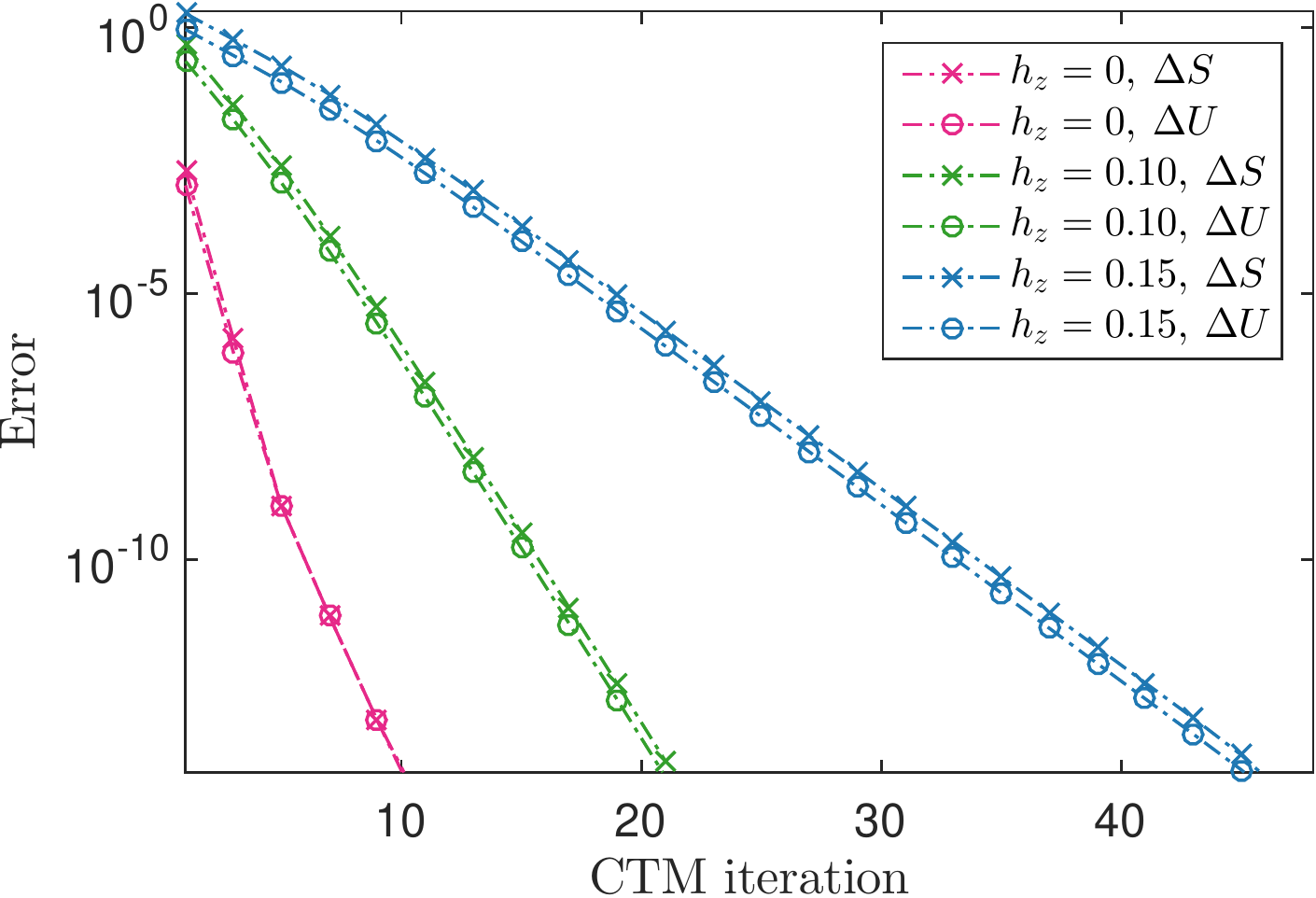}
\caption{Error in the modular matrices with respect to the exact results, $\Delta S = ||S-S_{exact}||$ and $\Delta U = ||U-U_{exact}||$, as a function of CTM iteration $N$, corresponding to a linear system  size $L=2N+2$, here for $D=2$. }
\label{Fig:ErrorSUmatrices}
\end{center}
\end{figure}

In the topological phase ($h_z = 0.1$) we can make similar observations as at the exactly solvable point. We again obtain tensors which, after a suitable gauge change, are approximately \ZZ symmetric, and after symmetrizing them we are able to successfully extract the correct TEE and modular matrices with a very high accuracy. Interestingly, after the gauge change the deviation from a \ZZ symmetric tensor is larger for $D=3$ than for $D=2$, probably because there is more freedom in these tensors to add off-diagonal elements which do not affect the energy in a significant way. We note also that the change in energy from $D=2$ to $D=3$ is very small here. In  Fig.~\ref{Fig:ErrorSUmatrices} we show the error in the modular matrices with respect to the exact results as a function of CTM iteration, showing that after sufficiently many steps (i.e. for sufficiently large system sizes depending on the correlation length in the system) the $U$ and $S$ matrices are accurately reproduced. 

Outside the topological phase ($h_z = 0.18$) we consistently find a vanishing TEE and a non-degenerate transfer matrix spectrum, which allows us to correctly identify the trivial phase. Since the ground state is no longer degenerate, the CTM approach only yields a state with a finite norm in the even-even sector, whereas the other sectors exhibit a vanishingly small norm, such that the (trivial) modular matrices cannot be computed in a meaningful way here. We further note that we do find a gauge in which the tensors are close to be \ZZ symmetric, however, the \ZZ virtual symmetry alone does not automatically imply a topologically ordered phase.

In Tab.~\ref{Tab:ComparissonD3} we present data of two additional examples for $D=3$ ($\chi=80$) in and outside the topological phase along the self-dual line, $(h_x,h_z) = (h,h)$ with $h=\{0.1,0.18\}$ for which we find similar results as in above case. (The critical point is located at $h_c=0.170(1)$~\cite{wu2012}).

\begin{table}[htb]
\begin{tabular} {c | c  c }
  
 &  \hsp  $h_z=h_z = 0.15$  \hsp &  \hsp $h_z=h_x = 0.20$   \hsp\\
\hline
$E$ &   \hsp$ -1.04952 $   &  \hsp $ -1.10492 $    \hsp\\ 
\hline
 $|E - E_{Z_2}|$   &  \hsp $ 2.1\e{-5} $   \hsp &  \hsp $ 5.1\e{-3} $   \hsp\\
\hline
 $\gamma -  \log{2}$   &  \hsp $ 2.2\e{-13}$   \hsp&   \hsp$ 0.6931$   \hsp\\
 \hline 
 $\delta_{Z_2} / ||A||^2$   &  \hsp $ 6.9\e{-4}$   \hsp&  \hsp $ 0.155 $   \hsp\\
%
%
\end{tabular}
\caption{Same as in Tab.~\ref{Tab:ComparissonD2}, here for $D=3$, $\chi=80$ for  fields along $h=h_z=h_x$. Also here the same $S$ and $U$ matrices are obtained in the topologically ordered phase as in Tab.~\ref{Tab:h0}, with an accuracy which is close to machine precision.
}
\label{Tab:ComparissonD3}
\end{table}

\subsection{Phase transition as a function of \texorpdfstring{$h_z$}{hz}}
\begin{figure}[tb]
\begin{center}
\includegraphics[width=\columnwidth,trim={0cm 0cm 0cm 0cm},clip]{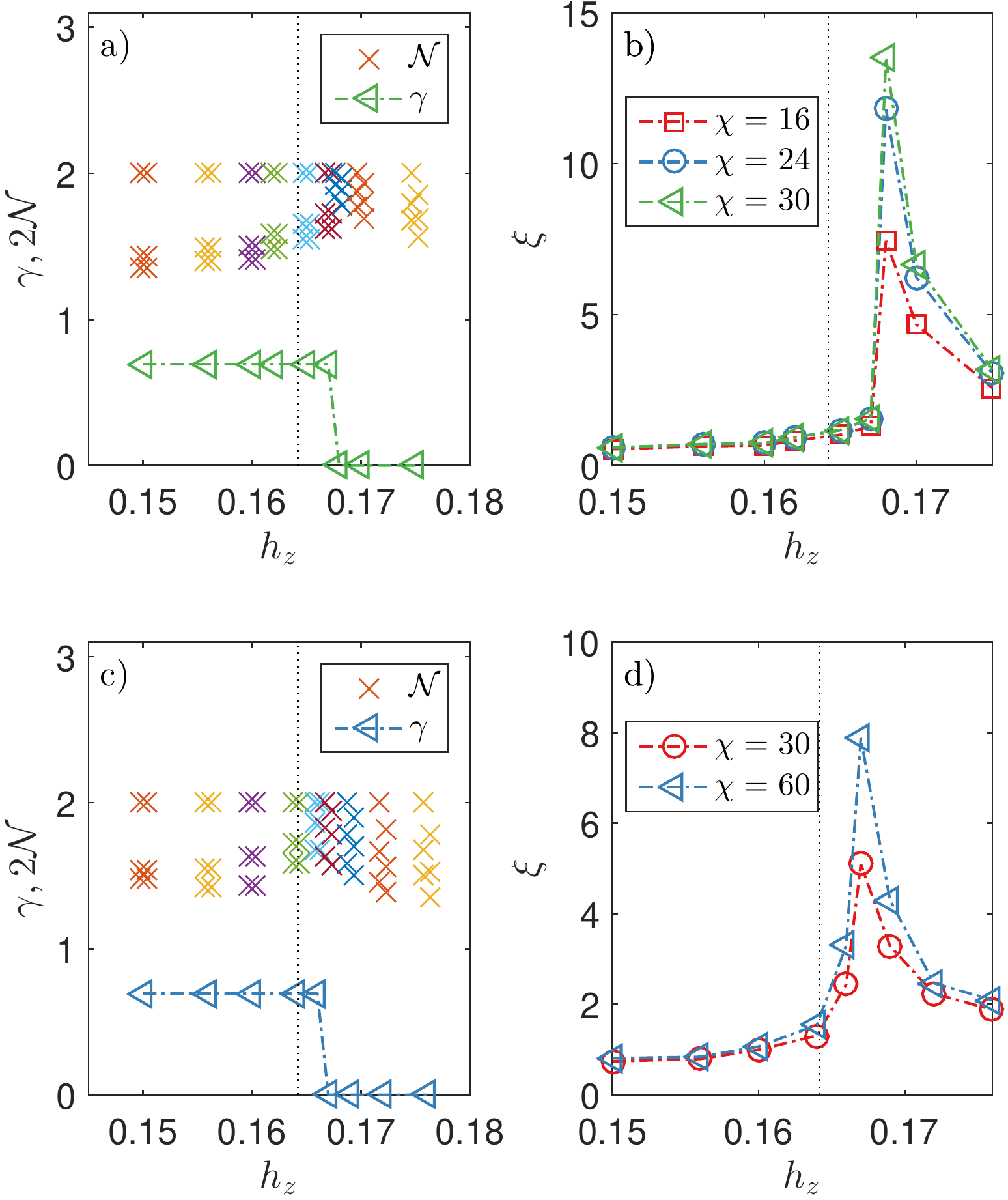}
\caption{
(a) The TEE $\gamma$ and the dominant eigenvalues of the transfer matrix $\mathcal{N}$ across the phase transition as a function of the magnetic field $h_z$, obtained for $D=2$ ($\chi=30$). The dashed line shows the Monte-Carlo result for the location of the phase transition~\cite{wu2012}. (b) Correlation length $\xi$ as a function of $h_z$ across the phase transition for different values of the boundary dimension $\chi$, exhibiting a peak at the $D=2$ critical point. (c-d) Same as in (a-b) for bond dimension $D=3$ ($\chi=60$ in (a)).
}
\label{Fig:OnlyHz}
\end{center}
\end{figure}

Finally we test the approach  for magnetic fields close to the phase transition as a function of $h_z$ (with $h_x=0$) to locate the critical point. In Fig.~\ref{Fig:OnlyHz}(a) we present the $D=2$ results for the TEE $\gamma$ (obtained from linear fits to $S_2(L)$ with $L$ between 200 and 400) which exhibits a clear jump from $\log(2)$ down to zero at $h_z=0.1675(5)$ which is close but not equal to the Monte-Carlo result $h_c=0.164237(2)$~\cite{wu2012} due to finite $D$ effects. In Fig.~\ref{Fig:OnlyHz}(a) we also show the largest few values of the transfer matrix spectrum $\mathcal{N}$ where we clearly find a degeneracy within the topologically ordered phase up to the $D=2$ critical point as expected. At the phase transition we consistently find a peak in the correlation length $\xi$ (Fig.~\ref{Fig:OnlyHz}(b)), computed from the two leading (non-degenerate) eigenvalues of the transfer matrix of the gauge-transformed state.
With increasing bond dimension the location of the phase transition approaches the Monte-Carlo result, as shown in Figs.~\ref{Fig:OnlyHz}(c-d) with a transition value  $h_z=0.1665(5)$ for $D=3$.

These results demonstrate that our  approach is applicable also close to a phase transition (with an accuracy on the location of the critical point depending on $D$ as in conventional phase transitions~\cite{corboz18,rader18}).


\section{Summary and discussion}
\label{Sec:summary}
In this work we have demonstrated that it is possible to correctly identify a topologically ordered phase using unbiased iPEPS simulations, i.e. where we start the optimization from random initial tensors without imposing the corresponding virtual symmetry on the tensors. As an example we considered the toric code model in a magnetic field where, within the topologically ordered phase, the tensors should exhibit a virtual \ZZ symmetry. 
We found  that, after a suitable gauge change, the resulting tensors are approximately \ZZ symmetric, and they can be fully symmetrized a posteriori to generate a stable topologically ordered state, exhibiting the correct topological entanglement entropy and modular S and U matrices.

What are the implications of our findings? So far, a common believe was that the  virtual symmetry needs to be imposed on the tensors in order to obtain and identify the correct phase, which is not a problem if the type of topologically order is known beforehand. However, if it is not known one would need to run a separate simulation for each possible virtual symmetry (each corresponding to another topologically ordered phase) which is not efficient since the optimization of the tensors is the computationally most  expensive part. In our approach, in contrast, the idea is to perform a single unbiased simulation using unconstrained tensors, and determine the (approximate) virtual symmetry a posteriori, as a part of the analysis of the state. This is computationally cheaper and more in the spirit of unbiased  tensor network calculations. We note that during completion of this work a different approach to study topological phases based on unbiased iPEPS simulations was presented in Ref.~\onlinecite{francuz19}.

In this paper we have also developed a variant of the CTM method where the parity on the boundary can be controlled, which is computationally more efficient than schemes based on TRG~\cite{he2014a} to compute the modular S and U matrices in the topological phase. Furthermore, we have introduced a gradient-based energy minimization algorithm based on a summation of tensor environments, which is simpler and more efficient than approaches based on a summation of Hamiltonian terms~\cite{corboz2016,vanderstraeten2016},  especially for models with interactions beyond nearest-neighbor sites and multi-site interactions. 

\begin{acknowledgements}
This project has received funding from the European Research Council (ERC) under the European Union's Horizon 2020 research and innovation programme (grant agreement No 677061). This work is part of the D-ITP consortium, a program of the Netherlands Organization for Scientific Research (NWO) that is funded by the Dutch Ministry of Education, Culture and Science~(OCW). 
\end{acknowledgements}

\bibliographystyle{apsrev4-1}
\bibliography{TCrefs}

\end{document}